\documentclass[12pt,a4paper]{article}
\usepackage[DIV=13]{typearea}
\usepackage[onehalfspacing]{setspace}
\usepackage[english]{babel}
\usepackage[ansinew]{inputenc}
\usepackage{amssymb,amsmath,bm,array,dsfont,graphicx,natbib}
\usepackage[colorlinks,citecolor=black,linkcolor=black,urlcolor=black]{hyperref}
\usepackage[hang,tight]{subfigure}
\usepackage{authblk}

\newcommand{\newoperator}[3]{\newcommand*{#1}{\mathop{#2}#3}}
\newcommand{\renewoperator}[3]{\renewcommand*{#1}{\mathop{#2}#3}}
\newcommand{\calI}{\mathcal{I}}

\newcommand{\mA}{A}

\newcommand{\mB}{B}

\newcommand{\vc}{c}

\newcommand{\vd}{d}

\newcommand{\ve}{e}
\newcommand{\mF}{F}

\newcommand{\mH}{H}
\newcommand{\vh}{h}
\newcommand{\mI}{I}

\newcommand{\mM}{M}

\newcommand{\mN}{N}

\newcommand{\mR}{R}

\newcommand{\vu}{u}

\newcommand{\vv}{v}

\newcommand{\mX}{X}
\newcommand{\vx}{x}

\newcommand{\vy}{y}
\newcommand{\mZ}{Z}
\newcommand{\vz}{z}
\newcommand{\valpha}{\alpha}
\newcommand{\vbeta}{\beta}

\newcommand{\vepsi}{\epsi}

\newcommand{\vzeta}{\zeta}

\newcommand{\vkappa}{\kappa}

\newcommand{\vnu}{\nu}
\newcommand{\vxi}{\xi}

\newcommand{\vtau}{\tau}

\newcommand{\vphi}{\phi}

\newcommand{\vomega}{\omega}
\newcommand{\mGamma}{\varGamma}

\newcommand{\mTheta}{\varTheta}
\newcommand{\mLambda}{\varLambda}

\newcommand{\mSigma}{\varSigma}

\newcommand{\mPhi}{\varPhi}

\renewoperator{\Re}{\mathrm{Re}}{\nolimits}
\renewoperator{\Im}{\mathrm{Im}}{\nolimits}
\makeatletter
\newcommand{\rd}{\@ifnextchar^{\DIfF}{\DIfF^{}}}
\def\DIfF^#1{%
   \mathop{\mathrm{\mathstrut d}}%
   \nolimits^{#1}\gobblespace}
\def\gobblespace{\futurelet\diffarg\opspace}
\def\opspace{%
   \let\DiffSpace\!%
   \ifx\diffarg(%
   \let\DiffSpace\relax
   \else
   \ifx\diffarg[%
   \let\DiffSpace\relax
   \else
   \ifx\diffarg\{%
   \let\DiffSpace\relax
   \fi\fi\fi\DiffSpace}

\newcommand{\E}{\operatorname{E}}

\newoperator{\ip}{\mathrm{int}}{\nolimits}

\newcommand{\tr}{\operatorname{tr}}

\newcommand{\epsi}{\varepsilon}

\newcommand{\vzeros}{0}
\newcommand{\mZeros}{0}
\newcommand{\mzeros}{\mZeros}

\newcommand{\beq}{\begin{equation}}
\newcommand{\eeq}{\end{equation}}
\newcommand{\bal}{\begin{align*}}
\newcommand{\eal}{\end{align*}}

\newcommand{\bmat}{\begin{bmatrix}}
\newcommand{\emat}{\end{bmatrix}}

\newcommand{\bsmat}{\begin{smallmatrix}}
\newcommand{\esmat}{\end{smallmatrix}}

\title{Multivariate Fractional Components Analysis}
\author[1,2]{Tobias Hartl}
\author[3]{Roland Weigand\footnote{Corresponding author. E-Mail: roland.weigand@posteo.de}}

\affil[1]{University of Regensburg, 93053 Regensburg, Germany}
\affil[2]{Institute for Employment Research (IAB), 90478 Nuremberg, Germany}
\affil[3]{AOK Bayern, 93055 Regensburg, Germany}
\date{January 2019}

\begin{document}
\maketitle

\thispagestyle{empty}
\setcounter{page}{0}

\paragraph{\bf Abstract.}

We propose a setup for fractionally cointegrated time series which is formulated in terms of latent integrated and short-memory components. It accommodates nonstationary processes with different fractional orders and cointegration of different strengths and is applicable in high-dimensional settings. In an application to realized covariance matrices, we find that orthogonal short- and long-memory components provide a reasonable fit and competitive out-of-sample performance compared to several competing 
methods.


\paragraph{\bf JEL-Classification.}

C32, C51, C53, C58.

\newpage

\section{Introduction}

Multivariate fractional integration and cointegration models have proven valuable in a wide range of empirical applications from macroeconomics and finance. They generalize the standard concept of cointegration by allowing for non-integer orders of integration both for the observations and for equilibrium errors; see \citet{GilHua2008} for a literature review. In the field of macroeconomics, such models have turned out to be relevant in analyses of purchasing power parity beginning with \citet{CheLai1993}, of the relation between unemployment and input prices \citep{CapGil2002} and of broader models for economic fluctuations \citep{Mor2006}. The empirical finance literature has considered fractional cointegration, e.g., for analysing international bond returns \citep{DueSta1998}, for modeling co-movements of stock return volatilities \citep{BelMor2006}, for assessing the link between realized and implied volatility \citep{Nie2007} and for quantifying risk in strategic asset allocation problems \citep{SchTscBu2008}. From a methodological point of view, semiparametric techniques for inference on the cointegration rank, the cointegration space and memory parameters have been very popular among empirical researchers, although the development of optimal parametric inferential methods for models with triangular or fractional vector error correction representations has recently made considerable progress \citep[see, e.g.,][]{RobHua2003,AvaVel2009,Las2010,JohNie2012}.

Despite their flexibility and their computationally simple treatment, semiparametric models are limited in scope since they aim to describe low-frequency properties only and are hence not appropriate for impulse response analysis and forecasting. While semiparametric techniques have been developed to cope with multivariate processes of different integration orders and multiple fractional cointegration relations of different strenghts \citep{CheHur2006,HuaRob2010,Hua2009}, there seems to be a lack of parametric models of such generality. Furthermore, the usual error correction and triangular models with their abundant parametrization are not deemed appropriate for time series of dimension, say, larger than five.

In this paper, we propose new models for multivariate fractionally integrated and cointegrated time series which are formulated in terms of latent purely fractional and additive short-memory components. With a ``type II'' definition of fractional integration \citep{Rob2005}, this approach allows for a flexible modeling of possibly nonstationary time series of different fractional integration orders. It permits cointegration relations of different strengths as well as polynomial cointegration \citep[multicointegration in the terminology of][]{GraLee1989}, i.e., cointegration between the levels of some time series and their (fractional) differences, and guarantees a clear representation of the long-run characteristics. Consequently, our model is among the most general setups regarding its integration and cointegration properties, compared to popular existing models for cointegrated processes. The unobserved components formulation benefits the modeling of relatively high-dimensional time series. For this situation we propose a parsimonious parametrization based on dimension reduction and dynamic orthogonal components in the spirit of \citet{PanYao2008} and \citet{MatTsa2011}. 

In contrast to our parametric approach, latent fractional components have mostly been studied in semiparametric frameworks. \citet{RayTsa2000} use semiparametric memory estimators and canonical correlations to infer the existence of common fractional components, \citet{Mor2004} proposes a frequency domain principal components estimator, \citet{Mor2007} estimates components of a single fractional integration order by univariate permanent-transitory (or persistent-transitory) decompositions followed by a principal component analysis of the permanent (or persistent) components and \citet{LucVer2015} estimate their fractional factor model by fitting long-memory models to the principal components of a large panel of time series. In a setup closest to ours, \citet{CheHur2006} suggest a semiparametric frequency domain methodology to identify and estimate cointegration subspaces which annihilate fractional components of different memory.

Recent parametric frameworks competing to ours have either been much more restrictive, or have a different focus, e.g., on numerical simulation-based estimation methods, or on panel data analysis. Using a Bayesian approach, \citet{HsuRayBr1998} discuss a bivariate process sharing one stationary long-memory component, while \citet{MesKooOo2016} consider simulated maximum likelihood estimation of models with one or more latent stationary ARFIMA components. On the other hand, \citet{ErgVel2017} as well as \citet{Erg2017} focus on the elimination of common fractional components and are motivated as alternatives to prior unit root testing. Contrary to our approach, they eliminate the common factor structure, which they treat as nuisance.


As the second main contribution of this paper, our model is applied to forecasting daily realized covariance matrices. In this setup, the strengths of our approach become apparent. In realized covariance modelling, typically high-dimensional processes with strong persistence and a pronounced co-movement in the low-frequency dynamics are considered. In time series of log variances and z-transformed correlations for six US stocks, we find that common orthogonal short- and long-memory components with two different fractional integration orders provide a reasonable fit. Since the dimension of the dataset is reduced to a smaller number of latent processes, our model becomes a factor model. A pseudo out-of-sample study shows that the fractional components model provides a superior forecasting accuracy compared to several competitor methods. In addition to the favorable forecast properties, our methods can be applied to study the cointegration properties of stock market volatilities. These are of particular importance for longer-term portfolio hedging and the analysis of systematic risk.  

The paper is organized as follows. Section \ref{sec:model} introduces the general setup and clarifies its integration and cointegration properties. In section \ref{sec:alternative}, its relation to existing models for multivariate integrated time series is discussed. In section \ref{sec:dimred}, a specific model appropriate for relatively high-dimensional processes is considered. The empirical application to realized covariance matrices and a pseudo out-of-sample assessment are contained in section \ref{sec:application} before section \ref{sec:conclusion} concludes.

\section{The general setup} \label{sec:model}

We consider a linear model for a $p$-dimensional observed time series $\vy_t$, which we label a \emph{fractional components} (FC) setup,
\beq \label{eq:FC}
\vy_t = \mLambda \vx_t + \vu_t, \qquad t=1,\ldots,n.
\eeq
The model is formulated in terms of the latent processes $\vx_t$ and $\vu_t$ where $\mLambda$ will always be assumed to have full column rank and the components of the $s$-dimensional $\vx_t$ are fractionally integrated noise according to
\beq \label{eq:FC_frac}
\Delta^{d_j} x_{jt} = \xi_{jt}, \qquad j=1,\ldots,s.
\eeq
In principle, $s>p$ is possible, but we only consider cases where $s \leq p$ here.
For a generic scalar $d$, the fractional difference operator is defined by
\beq \label{eq:fracdiff}
\Delta^d = (1-L)^d = \sum_{j=0}^{\infty} \pi_j(d)L^j, \quad \pi_0(d) =1, \quad \pi_j(d) =
\frac{j-1-d}{j}  \pi_{j-1}(d), \; j\geq 1,
\eeq
where $L$ denotes the lag or backshift operator, $Lx_t = x_{t-1}$. We adapt a nonstationary type II solution of these processes \citep{Rob2005} and hence treat $d_j\geq0.5$ alongside the asymptotically stationary case $d_j<0.5$ in a continuous setup, by 
setting starting values to zero, $x_{jt}=0$ for $t\leq 0$. Nonzero initial values have been considered for observed fractional processes by \citet{JohNie2012}, but are not straightforwardly handled for our unobserved processes. The solution is based on the truncated operator $\Delta_+^{-d_j}$ \citep{Joh2008} and given by
\[
x_{jt} = \Delta_+^{-d_j} \xi_{jt} = \sum_{i=0}^{t-1} \pi_i(-d_j) \xi_{j,t-i}, \qquad j=1,\ldots,s.
\]
Without loss of generality let the components be arranged such that $d_1 \geq \ldots \geq d_s$.

We assume $d_j>0$ for all $j$ in what follows, so that $x_t$ governs the long-term characteristics of the observations $\vy_t$. These are complemented by additive short-run dynamics which we describe by stationary vector ARMA specifications for $\vu_t$ in the general case. This ARMA process is given by
\beq \label{eq:FC_varma}
\mPhi(L) \vu_t = \mTheta(L) \ve_t, \qquad t=1,\ldots,n,
\eeq
where $\mPhi(L)$ and $\mTheta(L)$ are a stable vector autoregressive polynomial and an invertible moving average polynomial, respectively. The disturbances $\vxi_t$ and $\ve_t$ jointly follow a Gaussian white noise ($N\!I\!D$) sequence such that
\beq \label{eq:FC_noise}
\vxi_t \sim N\!I\!D (\vzeros, \mSigma_{\xi}), \quad
\ve_t  \sim N\!I\!D (\vzeros, \mSigma_{e})\quad \text{and} \quad \E(\vxi_t\ve_t')=\mSigma_{\xi e},
\eeq
where at this stage, before turning to identified and empirically relevant model specifications below, we do not consider restrictions on the joint covariance matrix, but only require $\mSigma_{\xi}$ to have strictly positive entries on the main diagonal.

Some remarks regarding the general FC setup are in order. The model as given in \eqref{eq:FC} is not identified without further restrictions on the loading matrix $\mLambda$, on the vector ARMA coefficients and on the noise covariance matrix. While restrictions on $\mSigma_{\xi}$ and $\mLambda$ may be based on results in dynamic factor analysis as will be seen below, choosing specific parametrizations for $\vu_t$ will depend on characteristics of the data and on the purpose of the empirical analysis. Identified vector ARMA structures like the echelon form \citep[see][chapter 12]{Luet2005} can be used for a rich parametrization, while a multivariate structural time series approach as described in \citet{Har1991} integrates nicely with the unobserved components framework considered in this paper and allows for more restricted parameterization, e.g., by individual or common stochastic cycle components. Below, we introduce a parsimonious model well-suited to relatively high dimensions which is conceptually based on dimension reduction and orthogonal components.

For a characterization of the integration and cointegration properties of our model, we adapt the definitions of these concepts from \citet{HuaRob2010}, which prove useful here. Hence, a generic scalar process $\rho_t$ is called integrated of order $\delta$ or $I(\delta)$ if it can be written as $\rho_t = \sum_{i=1}^l \Delta_+^{-\delta_i}\nu_{it}$, where $\delta = \max_{i=1,\ldots,l} \{ \delta_i\}$ and $\vnu_t=(\nu_{1t},\ldots,\nu_{lt})'$ is a finite-dimensional covariance stationary process with spectral density matrix which is continuous and nonsingular at all frequencies. A vector process $\vtau_t$ is called $I(\delta)$ if $\delta$ is the maximum integration order of its components. We call the process $\vtau_t$ cointegrated if there exists a nonzero vector $\vbeta$ such that $\vbeta'\vtau_t$ is $I(\gamma)$ where $\delta-\gamma>0$ will be referred to as the strength of the cointegration relation. The number of linearly independent cointegration relations with possibly differing $\gamma$ is called cointegration rank of $\vtau_t$.

By these definitions, $x_{jt}$ is clearly $I(d_j)$ while both $\vx_t$ and $\vy_t$ are integrated of order $d_1$. We observe at least two different integration orders in the individual series of $\vy_t$ whenever $\mLambda_{i1}=0$ for some $i$ and $d_1>d_2$. More generally, $y_{it} \sim I(d_j)$, if $\mLambda_{i1} = \ldots = \mLambda_{i,j-1}=0$ but $\mLambda_{ij}\neq 0$.

To state the cointegration properties of the FC setup \eqref{eq:FC}, we assume that $s \leq p$, so that all fractional components are reflected by the integration and cointegration structure of $\vy_t$ and that $\mSigma_{\xi}$ is nonsingular. It is useful to identify all $q$ groups of $x_{jt}$ with identical integration orders and denote their respective sizes by $s_1$, \ldots, $s_q$, such that $d_{s_1+\ldots+s_{j-1}+1}= \ldots = d_{s_1+\ldots+s_j}$ and $s=\sum_{j=1}^q s_j$. Of course, if $q=s$, then $s_1 = \ldots = s_q=1$ and all components of $\vx_t$ have mutually different integration orders, while for $q=1$ it holds that $s=s_1$ and we observe $d_1=\ldots=d_s$.

To keep notation simple, for a generic matrix $\mA$ for which a specific grouping of rows and columns is clear from the context, we denote by $\mA^{(i,j)}$ the block from intersecting the $i$-th group of rows with the $j$-th group of columns.
A stacking of several groups of rows $i,\ldots,j$ and columns $k,\ldots,l$ is indicated by $\mA^{(i:j,k:l)}$. For a grouping in only one dimension we write $\mA^{(i)}$ or $\mA^{(i:j)}$, where it shall be clear from the context whether a grouping of rows or columns is considered. Furthermore, we denote the column space of a generic $k \times l$ matrix $\mA$ by $sp(\mA)\subseteq \mathds{R}^k$ and its orthogonal complement by $sp^{\bot}(\mA)$. Further, for $k > l$, the $k \times (k-l)$ orthogonal complement of $\mA$ will be denoted by $\mA_{\bot}$, which spans the $(k-l)$-dimensional space $sp^{\bot}(\mA)$.

According to the grouping of equal individual integration orders in $\vx_t$, we may therefore rewrite the FC process \eqref{eq:FC} as
\[
\vy_t = \mLambda^{(1)} \vx_t^{(1)} + \ldots + \mLambda^{(q)} \vx_t^{(q)} + \vu_t.
\]
Here, $\mLambda^{(j)}$ is a $p \times s_{j}$ submatrix of $\mLambda$ consisting of columns $\mLambda_{\cdot i}$ for which $s_1+\ldots+s_{j-1}<i\leq s_1+\ldots+s_{j}$, and $\vx_t^{(j)}$ is a $s_j$-dimensional subprocess of $\vx_t$ corresponding to components with memory parameter $d^{(j)}:=d_{s_1+\ldots+s_{j-1}+1}=\ldots = d_{s_1+\ldots+s_j}$. Whenever $s_1<p$, there exist $p-s_1$ linearly independent linear combinations $\vbeta_i'\vy_t \sim I(\gamma_i)$ and $\gamma_i<d_1$, so that fractional cointegration occurs. Due to our definition of cointegration, this may be a trivial case where a single component $y_{it}$ with integration order smaller than $d_1$ is selected. Since
\[
\mLambda^{(1)'}_{\bot} \vy_t = \mLambda^{(1)'}_{\bot}\mLambda^{(2)} \vx_t^{(2)} + \ldots + \mLambda^{(1)'}_{\bot}\mLambda^{(q)} \vx_t^{(q)} +\mLambda^{(1)'}_{\bot} \vu_t
\]
is integrated of order $d^{(2)}$, the columns of $\mLambda^{(1)}_{\bot}$ qualify as cointegration vectors and ${\cal S}^{(1)}:=sp^{\bot}(\mLambda^{(1)})$ is the $(p-s_1)$-dimensional cointegration space of $\vy_t$.

Whenever $s_1+s_2<p$, there are subspaces of ${\cal S}^{(1)}$ forcing a stronger reduction in integration orders. More generally, it holds that $\mLambda_{\bot}^{(1:j)'}\vy_t \sim I(d^{(j+1)})$ whenever $\sum_{i=1}^js_i<p$ and where we set $d^{(j+1)}=0$ for $j>s$. Analogously to \citet{HuaRob2010}, for $s=p$ and $j=1,\ldots,q-1$, we call ${\cal S}^{(j)}:=sp^{\bot}(\mLambda^{(1:j)})$ the $j$-th cointegration subspace of $\vy_t$, for which ${\cal S}^{(q-1)}\subset \ldots \subset {\cal S}^{(1)}$. For $p>s$, ${\cal S}^{(q)}\subset{\cal S}^{(q-1)}$ is a further such subspace. Cointegration vectors in ${\cal S}^{(q)}$ cancel all fractional components and hence reduce the integration order from $d_1$ to zero, the strongest reduction possible in our setup.

Besides this general pattern of cointegration relations, our model features an interesting special case with so-called polynomial cointegration, that is, cointegration relations where lagged observations nontrivially enter a cointegration relation. To see this possibility, consider a bivariate example similar to \citet{GraLee1989}, where $q=p=2$ and $\xi_{1t}=\xi_{2t}$, so that $\mSigma_{\xi}$ is singular and $x_{2t} = \Delta^{d_1-d_2} x_{1t}$. Augmenting the variables by a fractional difference as $\tilde{\vy}_t:=(y_{1t}, y_{2t}, \Delta^{d_1-d_2}y_{2t})'$, we obtain a three-dimensional system where levels of $\vy_t$ enter a nontrivial cointegration relation with a fractional difference to achieve a reduction in integration order from $d_1$ to $\max\{2d_2-d_1,0\}<d_2$. 
Hence, our setup complements the model of \citet[][section 4]{Joh2008}, which was the first to handle polynomial cointegration in a fractional setup, and the results of \citet{CarSan2018}, who derive a Granger representation for the fractional VECM of \cite{Gra1986} under polynomial cointegration.
%

\section{Relations to other cointegration models} \label{sec:alternative}

In this section, we clarify the relation of the fractional components model \eqref{eq:FC} to popular existing representations for cointegrated processes and show how our model can be represented in alternative ways brought forward in the literature. While our model is among the most general setups with respect to its integration and cointegration properties, the additive modeling of short-run dynamics is new to the literature and gives rise to distinct parametrizations not possible within other representations in a similarly convenient way.

\paragraph{Error correction models.} The most popular representation of cointegrated systems in the $I(1)$ setting is the vector error correction form. Since an early mention by \citet{Gra1986}, in the fractionally integrated case, e.g., \citet{AvaVel2009}, \citet{Las2010} and \citet{JohNie2012} have recently considered such models. In terms of the integration and cointegration properties, the fractional error correction setups are typically restricted to the special case with $q=2$ and $s=p$, such that the observed variables are integrated of order $d^{(1)}$ and there exist $p-s_1$ cointegration relations with errors of order $d^{(2)}$.

Defining the fractional lag operator $L_b:=1-\Delta^b$ \citep{Joh2008}, we are able to derive the error correction representation for this special case of our model; see appendix \ref{app:a1}. It is given by
\beq \label{eq:vecm}
\Delta^{d^{(1)}} \vy_t =  \valpha \vbeta'  L_{d^{(1)}-d^{(2)}}\Delta^{d^{(2)}} \vy_t + \vkappa_t,
\eeq
where we find $\valpha \vbeta'=-\mLambda^{(2)}(\mLambda^{(1)'}_{\bot} \mLambda^{(2)})^{-1}\mLambda^{(1)'}_{\bot}$ to precede the error correction term, while
\[
\vkappa_t:=\mM(\mLambda^{(1)}\vxi_t^{(1)} + \Delta^{d^{(1)}} \vu_t )-\valpha \vbeta'(\mLambda^{(2)}\vxi_t^{(2)} + \Delta^{d^{(2)}} \vu_t)
\]
is integrated of order zero and $\mM$ is defined in \eqref{eq:defm}.

The model differs both from the models of \citet{AvaVel2009} and from the representation of \citet{Joh2008} in the way short-run dynamics are modeled. The literature has considered (fractional) lags of differenced variables and possibly of error correction terms in the VECM representation. Our setup, in contrast, generates autocorrelated $\vkappa_t$ by filtering the latent $\vu_t$ with fractional difference operators. Hence, adding lags of $\Delta^{d^{(1)}} \vy_t$ in the model \eqref{eq:vecm} is only an approximate solution and achieving a desired approximation quality may require estimating a large number of parameters.

As we have discussed above, \citet{Joh2008} proposes a polynomially cointegrated generalization of his $\mathrm{VAR}_{d,b}$ model which allows terms integrated of orders $d$, $d-b$ and $d-2b$ in the Granger representation \citep[][theorem 9]{Joh2008}. Even compared to that specification, our model allows for more general patterns of integration orders and cointegration strengths, since we only assume $d_j > 0$ for all $j$. More in line with the generality envisaged in this paper, \citet[][equation 14]{TscWebWe2013a} present a model with error correction term and different integration orders, while \citet{LaVel2014} sequentially fit error correction models to test for cointegration relations of possibly different strengths.

\paragraph{Vector ARFIMA.} An interesting special case of \eqref{eq:FC} occurs for $s=p$ and $\mLambda=\mI$, where each series in $y_{it}$ is driven by a single fractional component and $y_{it} \sim I(d_i)$. This resembles standard vector ARFIMA models with possibly different integration orders; see, e.g., \citet{Lob1997} who labels the popularly termed vector ARFIMA class considered here as ``model A''. A frequently used submodel is the fractionally integrated vector autoregressive model discussed by \citet{Nie2004b}. The main difference to these approaches is our additive modeling of short-run dynamics, whereas in the vector ARFIMA setup weakly dependent vector ARMA instead of white noise processes are passed through the fractional integration filters.

Our model belongs to the class of vector ARFIMA processes for integer $d_j\in \{1,2,\ldots\}$, but not for general fractional integration orders. For the case of integer $d_j$, note that $(\vx_t',\vu_t')'$ is a finite-order vector ARMA process, and hence $\vy_t$ as a linear combination is itself in the ARMA class; see \citet{Luet1984}. For general vector ARFIMA processes, a similar conclusion does not hold. To see this, consider a stylized univariate case of our model with $p=s=1$, where $\Delta^d x_t =\xi_t$ and $(1- \phi L) u_t = e_t$. First note that $(\Delta^d x_t, u_t)'$ has an ARMA structure, and hence $(x_t, u_t)'$ is a vector ARFIMA process. Expanding $(1-\phi L) \Delta^d x_t = (1-\phi L)\xi_t$ and $(1-\phi L) \Delta^d u_t = \Delta^de_t$, we can write the sum, belonging to the fractional components model class, as
\beq \label{eq:noARFIMA}
(1-\phi L) \Delta^d y_t = (1-\phi L) \xi_t + \Delta^d e_t.
\eeq
The right hand side of this expression is not a finite-order MA process in general, as it has nonzero autocorrelations for all lags, and hence, the process does not belong to the ARFIMA class for non-integer $d$. 

\paragraph{Triangular representations.} The models discussed so far have restricted integration or cointegration properties as compared to our model. Even in the most general setup of \cite{Joh2008}, the integration orders are restricted to be $d$, $d-b$, $d-2b$ for polynomial cointegration. In contrast, \citet{Hua2009} and \citet{HuaRob2010} have proposed a very flexible model which adapts the triangular form of \citet{Phi1991} and its generalization to processes with multiple unit roots \citep{StoWat1993} to the fractional cointegration setup.

To derive the triangular representation for our model, we assume that the variables in $\vy_t$ are ordered in a way that $\mLambda^{(1:j,1:j)}$ is nonsingular for $j=1,\ldots,q$
and restrict attention to the case $s=p$ for notational convenience. The variables are partitioned according to the groups of different integration orders in $\vx_t$ as $\vy_t^{(j)}:=(y_{s_1+\ldots+s_{j-1}+1}, \ldots, y_{s_1+\ldots+s_j})'$, $j=1,\ldots,q$. The first block in the triangular system is
\beq \label{eq:tria1}
\Delta^{d_1} \vy^{(1)}_t = \mLambda^{(1,1)} \vxi_{t}^{(1)} +
\mLambda^{(1,2)} \Delta^{d_1-d_2} \vxi_{t}^{(2)}+\ldots +\mLambda^{(1,q)} \Delta^{d_1-d_q} \vxi_{t}^{(q)}+ \Delta^{d_1} \vu_t \quad (:= \vomega_t^{(1)}),
\eeq
where $\vomega_t^{(1)}$ is integrated of order zero. The general expression for the $j$-th block of the triangular system is derived in appendix \ref{app:a1} for $j=2,\ldots,q$, and given by
\begin{align}
\Delta^{d^{(j)}} \vy^{(j)}_t &= \mLambda^{(j,1:(j-1))} (\mLambda^{(1:(j-1),1:(j-1))})^{-1}\Delta^{d^{(j)}} \vy^{(1:(j-1))}_t +\vomega_t^{(j)} \label{eq:triaj}\\
&= -\mB^{(j,1)} \Delta^{d^{(j)}}\vy^{(1)}_t - \ldots - \mB^{(j,j-1)} \Delta^{d^{(j)}}\vy^{(j-1)}_t + \vomega_t^{(j)}, \nonumber
\end{align}
where also $\vomega_t^{(j)}$ is integrated of order zero for $j=2,\ldots,q$. By inverting the fractional difference operators we obtain
\beq \label{eq:triaB}
\mB \vy_t = (\Delta_+^{-d_1}\vomega_t^{(1)'}, \ldots,\Delta_+^{-d_q}\vomega_t^{(q)'})',
\eeq
where $\mB$ has a block triangular structure such that $\mB^{(i,i)}=\mI$ and $\mB^{(i,j)}=\mzeros$ for $i<j$. A re-ordering of the variables in $\vy_t$ yields the representation of \citet{HuaRob2010}.

This representation allows for a semiparametric cointegration analysis of our model using the methods of \citet{Hua2009} and \citet{HuaRob2010}. However, our model differs significantly from straightforward parametrizations of the triangular system, e.g., from assuming a vector ARMA process for $\vomega_t$, since in our setup $\omega_t$ as stated in \eqref{eq:omega} generally contains fractional differences that cannot be represented within the ARMA framework.

\paragraph{State space approaches.}

\citet{BauWag2012} have presented a state space canonical form for multiple frequency unit root processes of different (integer-valued) integration orders. Their discussion is based on unit root vector ARMA models which are separated in pure unit root structures and short-term dynamics. Although the analogy to our model is striking, there are notable differences between their unit root and our fractional setup. Firstly, as discussed in the paragraph on vector ARFIMA models (see \eqref{eq:noARFIMA}), the fractional components setup \eqref{eq:FC} is not nested within a general class comparable to the vector ARMA models, which form the basis of the discussion in \citet{BauWag2012}. Secondly, in their setting, the introduction of different integration orders is achieved by repeated summation of lower order integrated processes which themselves enter the observations to achieve polynomial cointegration. This is in contrast to the continuous treatment of integration orders in our (type II) fractional setup.

However, fractional components models could be constructed to straightforwardly extend the setup of \citet{BauWag2012}. Using the fractional lag operator $L_b=1-\Delta^b$ instead of $L$ in the short-run dynamic specification \eqref{eq:FC_varma}, a stable vector ARMA$_b$ process can be defined by $\tilde{\mPhi}(L_b)\tilde{\vu}_t = \tilde{\mTheta}(L_b) \ve_t$ under suitable stability conditions \citep[][corollary 6]{Joh2008}. Then, replacing $\vu_t$ by $\tilde{\vu}_t$ in the model setup \eqref{eq:FC} with $d_j$ restricted to some multiple of $b$ ($d_j=i_jb$, $i_j \in \{1,2,\ldots\}$), the process $\vy_t$ is in the class of vector ARMA$_b$ models itself, while unit roots in the vector autoregressive polynomial generate the fractional $I(d_j)$ processes. Such a framework could be treated analogously to \citet{BauWag2012}, but the restriction that all integration orders are multiples of $b$ makes such a framework somewhat less flexible than ours.

\section{A dimension-reduced orthogonal components specification} \label{sec:dimred}

So far, we have considered a general modeling setup and discussed its integration and cointegration properties as well as its relation to existing approaches in the literature. We now turn to the discussion of a specific model from this class which bears potential for parsimonious modeling of long- and short-run dynamics in relatively high-dimensional applications. Besides its general interest, this will be the workhorse specification for the empirical application to realized covariance modeling in section \ref{sec:application}.

To introduce the model and emphasize its restrictions as compared to \eqref{eq:FC}, we decompose the short-term dependent process $\vu_t$ into an autocorrelated component, $\mGamma \vz_t$, where $\vz_t$ is a vector of $s_0$ mutually uncorrelated components with $s+s_0 \leq p$, and a Gaussian white noise component $\vepsi_t$, respectively. We label the result the \emph{dynamic orthogonal fractional components} (DOFC) model,
\beq \label{eq:DOC_FC}
\vy_t = \mLambda^{(1)} \vx_t^{(1)} + \ldots + \mLambda^{(q)} \vx_t^{(q)} + \mGamma \vz_t + \vepsi_t,
\eeq
where $\vx_t$ is generated by purely fractional processes \eqref{eq:FC_frac} as above, while
\[
(1-\phi_{j1}L -\ldots - \phi_{jk}L^k) z_{jt} = \zeta_{jt}, \qquad j=1,\ldots,s_0,
\]
are $s_0$ univariate stationary autoregressive processes of order $k$. Regarding the noise processes $\vxi_t$, $\vzeta_t$ and $\vepsi_t$, we assume mutual independence over leads and lags,
\[
\vxi_t \sim N\!I\!D (\mzeros, \mI),\qquad \vzeta_t \sim N\!I\!D (\mzeros, \mI)\qquad \text{and} \quad \vepsi_t \sim N\!I\!D (\mzeros, \mH),
\]
where $\mH$ is diagonal with entries $h_{i}>0$, $i=1,\ldots,p$. Note that for $s + s_0 < p$ the DOFC model is a factor model as it allows for dimension reduction. 

The model as specified in \eqref{eq:DOC_FC} and below is not identifiable without further information. Considering $\tilde{\vy}_t:= \Delta^{d^{(1)}} \vy_t$ instead of $\vy_t$ to meet the assumptions of \citet{HeaSol2004}, their theorem 4 suggests that groups of common components $\Delta^{d^{(1)}}\vx_t^{(1)}$, \ldots, $\Delta^{d^{(1)}}\vx_t^{(q)}$, $\Delta^{d^{(1)}}\vz_t$ can be disentangled (up to rotations within these groups) through their different shapes in spectral densities whenever $d^{(1)}>\ldots>d^{(q)}>0$. Still, there exist observationally equivalent structures with $\tilde{\mLambda}^{(j)} = \mLambda^{(j)} \mM^{-1}$ and $\tilde{\vx}_t^{(j)}=\mM \vx_t^{(j)}$ which satisfy the model restrictions for orthonormal $\mM$. Hence, we impose further restrictions on the loading matrices. As is standard practice in dynamic factor analysis, we set the upper triangular elements to zero such that $\mLambda_{rl}^{(j)}=0$ for $r<l$, $j=1,\ldots,q$, and $\mGamma_{rl}=0$ for $r<l$. Certain observables are thus assumed not to be influenced by certain factors.

The model \eqref{eq:DOC_FC} is very parsimonious considering that it includes both a rich fractional structure as well as short-run dynamics with co-dependence. This is possible by comprising three components of parsimony which have been brought forward in the statistical time series literature. Firstly, there are $p-s-s_0\geq 0$ white noise linear combinations of $\vy_t$. A strict inequality implies a reduced dimension in the dynamics of $\vy_t$ which is characteristic for so-called statistical factor models; see \citet{PanYao2008}, \citet{LamYaoBa2011} and \citet{LamYao2012}. In contrast, the model \eqref{eq:FC} does not belong to this class in general, since it allows for $s \geq p$ and general forms of autocorrelation in $\vu_t$. Secondly, all cross-sectional correlation stems from the common components which is a familiar feature from classical factor analysis \citep{AndRub1956}. Thirdly, both the fractional and the nonfractional components are mutually orthogonal for all leads and lags. 

Combined with semiparametric techniques of fractional integration and cointegration analysis, existing methods for statistical factor and dynamic orthogonal components analysis \citep{MatTsa2011} can be used to justify the model assumptions and may be useful in the course of model specification. For final model inference, maximum likelihood estimation based on a state space representation is the preferred method. Both steps  will be illustrated in the empirical application of the next section.

\section{An application to realized covariance modeling} \label{sec:application}

We apply the fractional components approach to the modeling and forecasting of multivariate realized stock market volatility which has recently received considerable interest in the financial econometrics literature.

\subsection{Data and recent approaches}

We use the dataset of \citet{ChiVoe2011} which comprises realized variances and covariances from six US stocks, namely (1) American Express Inc., (2) Citigroup, (3) General Electric, (4) Home Depot Inc., (5) International Business Machines and (6) JPMorgan Chase \& Co for the period from 2000-01-01 to 2008-07-30 ($n=2156$). The data are available from \url{http://qed.econ.queensu.ca/jae/2011-v26.6/chiriac-voev}.

\begin{figure}[h!]
\begin{center}\vspace{-0.5 cm}
\includegraphics[width=\textwidth]{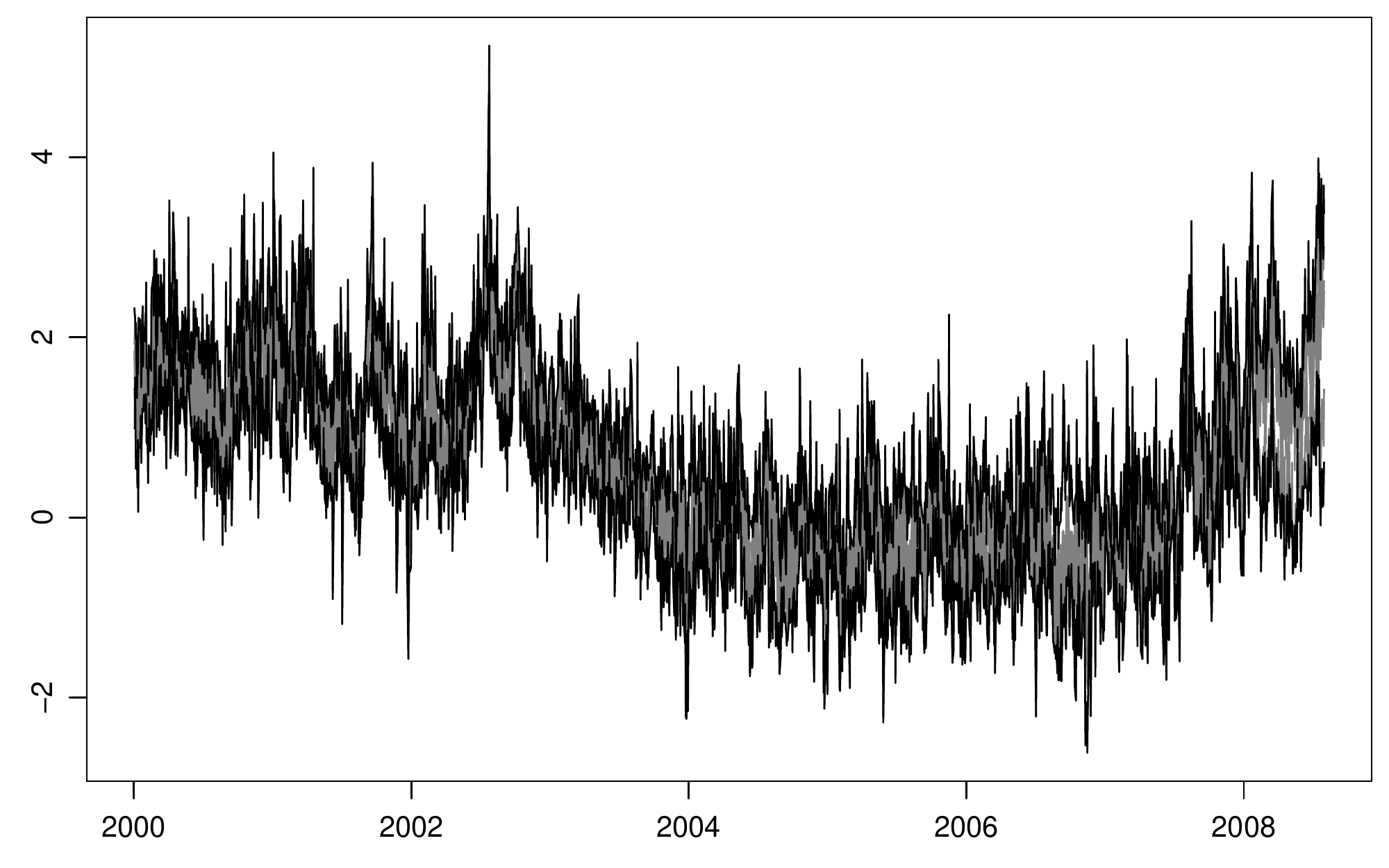}
\end{center}\vspace{-1cm} \caption{Time series plots of realized variances for the dataset described in section \ref{sec:application} (grey) together with maximum and minimum for all periods (black).}\label{fig:ts_var}
\end{figure}

\begin{figure}[h!]
\begin{center}\vspace{-0.5 cm}
\includegraphics[width=\textwidth]{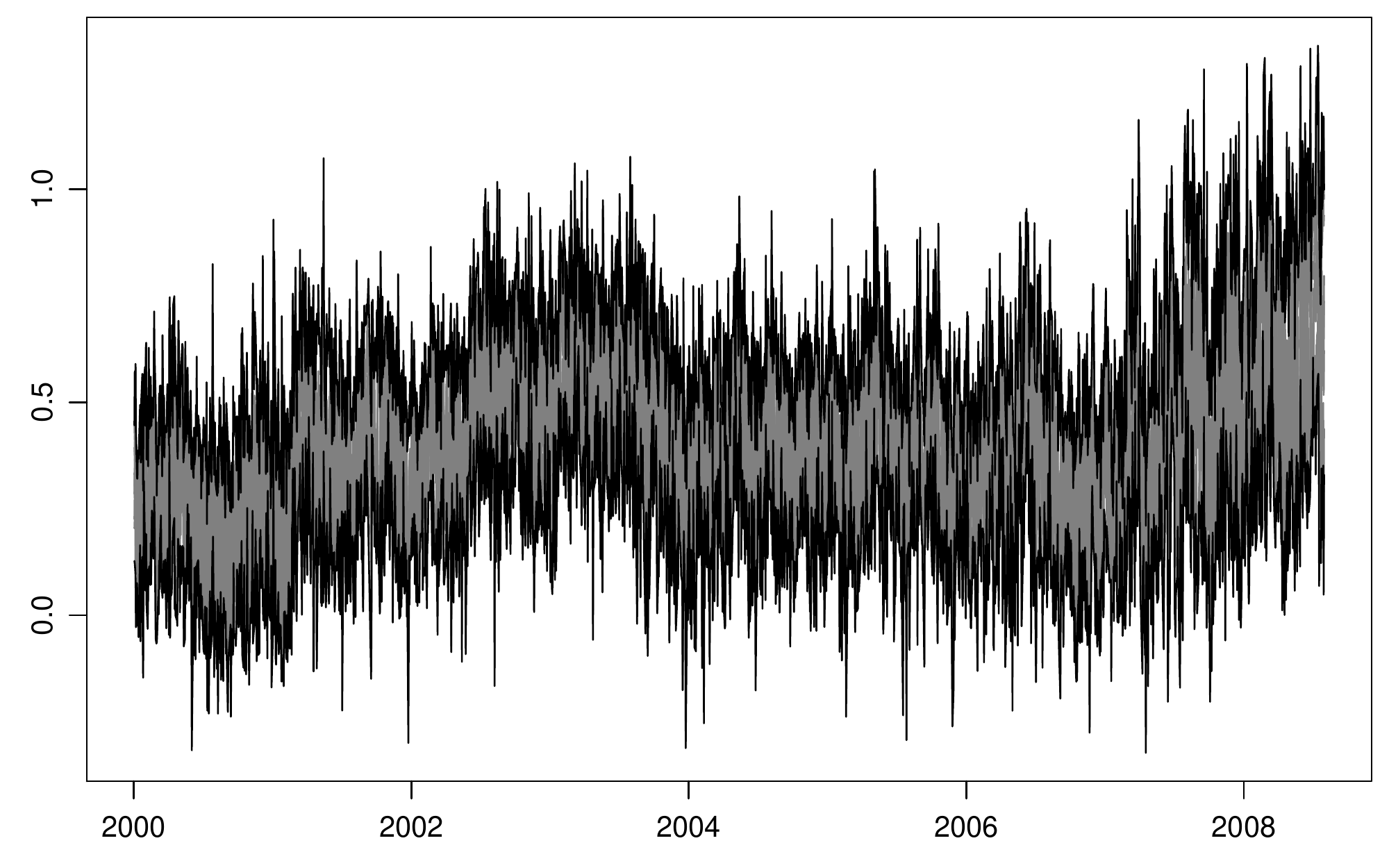}
\end{center}\vspace{-1cm} \caption{Time series plots of z-transformed realized correlations for the dataset described in section \ref{sec:application} (grey) together with maximum and minimum for all periods (black).}\label{fig:ts_cor}
\end{figure}

Different transformations of the realized covariance matrices have been applied to fit dynamic models to data of this kind. \citet{Wei2014} discusses these transforms and considers a general framework nesting several previously applied approaches. His results suggest that applying linear models to a multivariate time series of log realized variances along with z-transformed realized correlations is a reasonable choice in practice. We follow this approach and base our empirical study on the 21-dimensional time series
\beq \label{eq:logz}
\vy_t =  (\log(\mX_{11,t}), \ldots, \log(\mX_{66,t}), \mZ_{21,t}, \mZ_{31,t}, \ldots, \mZ_{65,t})',
\eeq
where $\mX_t$ is the $6 \times 6$ realized covariance matrix at period $t$, and the z-transforms are
\[
\mZ_{ij,t} = 0.5[\log(1+\mR_{ij,t})-\log(1-\mR_{ij,t})], \quad \mR_{ij,t}=\frac{\mX_{ij,t}}{\sqrt{\mX_{ii,t}\mX_{jj,t}}}.
\]
All time series (grey) of log variances and their maxima and minima for a given day $t$ (black) are depicted in figure \ref{fig:ts_var}, while z-transformed correlations are shown in figure \ref{fig:ts_cor}.

Recent approaches to modeling realized covariance matrices have successfully used long-memory specifications \citep{ChiVoe2011}, or found co-movements between the processes well-re\-pre\-sen\-ted by dynamic factor structures; see \citet{BauVor2011} and \citet{Gri2013}. In the related problem of forecasting univariate realized variances, factor models with long-memory dynamics have already been proposed. While \citet{BelMor2006} use frequency-domain principal components techniques to assess the low-frequency co-movements, \citet{LucVer2015} apply time-domain principal components to their high-dimensional series and apply fractional integration techniques to both estimated factors and idiosyncratic components. Recently, \citet{AsaMcA2014} have considered long-memory factor dynamics also for the modeling of realized covariance matrices, where again a semiparametric factor approach precedes a long-memory analysis in their two-step approach.

Our fractional components model DOFC \eqref{eq:DOC_FC}, applied to the time series \eqref{eq:logz}, offers various advantages to researchers and practitioners in the field.
(a) Our methods offer new insights in the integration and cointegration properties of stock market volatilities, for which fractional components structures of different integration orders have not been investigated so far. (b) Fractional cointegration between variances and correlations is of particular interest for the understanding of longer-term portfolio hedging and systemic risk assessment, but has not found attention in the existing literature. (c) Our state space approach for variances and correlations also features other relevant aspects of volatility modeling. It offers a separation into short-term and long-term components in the spirit of \citet{EngLee1999}, directly accounts for measurement noise, and is applicable in datasets of higher dimensions. The parameter-driven state space approach our specification enables yields (d) practicability in case of missing values, while it (e) straightforwardly carries over to stochastic volatility frameworks for daily return data in the spirit of \citet{HarRuiSh1994}.

\subsection{Preliminary analysis and model specification} \label{sec:semipar}

We investigate whether the constraints imposed in the DOFC model \eqref{eq:DOC_FC} are reasonable for the dataset under investigation. Semiparametric methods are used to assess these restrictions and to obtain reasonable starting values for the parametric estimation of our model.

The model \eqref{eq:DOC_FC} implies that there are $s+s_0$ components which govern the dynamics of $\vy_t$, and hence, for $p>s+s_0$, there is a dimension reduction in terms of the autocorrelation characteristics. \citet{PanYao2008} study time series with such properties and propose a sequential test to infer the dynamic dimension of the process, allowing for nonstationarity of the autocorrelated components. The algorithm sequentially finds the least serially correlated linear combinations of $\vy_t$, subsequently testing the null of no autocorrelation of these linear combinations. We apply 3 lags when detecting autocorrelations in what follows.

Applying this approach to our dataset,
we do not reject the null for eight linear combinations which can hence be treated as white noise. For the ninth such combination, the p-value for the multivariate Ljung-Box test drops from 0.1935 to 0.0002, so that the white noise hypothesis is rejected for reasonable significance levels. We conclude that there are $s+s_0=21-8=13$ components which account for the dynamic properties of the process. \citet{PanYao2008} also propose an estimator for the space of dynamic components $(\vx_t', \vz_t')'$. We call these estimates (rotated by principal components) \emph{the factors} in what follows.

Our model implies that $(\vx_t', \vz_t')'$ and hence a suitable rotation of the factors can be modelled as $s+s_0$ univariate time series which are mutually orthogonal at all leads and lags. This corresponds to the notion of dynamic orthogonal components as introduced by \citet{MatTsa2011} who provide methods to test for the presence of such a structure and to estimate the appropriate rotation. Using first differences of the factors to achieve stationarity as required by \citet{MatTsa2011} for suitable values of $d_j$, we find highly significant cross-correlations of the raw factors (the test statistic takes the value 4198.94 for a level 0.01 critical value of 625.80) while a dynamic orthogonal structure is not rejected for the rotated series, with a test statistic of 445.55 and a corresponding p-value close to one. The test result also holds if the test is conducted in levels. In what follows, the dynamic orthogonal components are computed from the factors in levels which slightly outperforms the difference-approach in simulations with fractional processes.\footnote{Results are available from the authors upon request.}

Due to their dynamic orthogonality, the rotation of \citet{MatTsa2011} identifies the single processes in $(\vx_t', \vz_t')'$ up to scale, sign and order. A preliminary analysis of the integration orders of $\vx_t$ can hence be undergone by a univariate treatment of these series. We investigate these integration orders by the exact local Whittle estimator allowing for an unknown mean \citep{Shi2010a}.

A possible grouping of components with equal integration orders is assessed by the methods proposed by \citet{RobYaj2002}, with the modifications for possibly nonstationary integration orders by \citet{NieShi2007}. The specific-to-general approach of \citet{RobYaj2002} sequentially tests for existence of $j=1,2,\ldots$ groups of equal integration orders. The sequence is terminated if for some $j^*$ there is a grouping for which within-group equality is not rejected, and for $j^*>1$ the grouping with highest p-value is selected. In our application, we restrict attention to possible groupings where, for $\hat{d}_{i_1}>\hat{d}_{i_2}>\hat{d}_{i_3}$, there is no group including both $i_1$ and $i_3$ but not $i_2$. For the tests of equal integration orders within the sequential approach, we consider the Wald test proposed by \citet{NieShi2007}, jointly testing all hypothesized equalities for a given grouping. We choose $m=\lfloor n^{0.5} \rfloor = 46$ as bandwidth and set the trimming parameter $h$ to zero, since the dynamic orthogonal components structure does not permit fractional cointegration.

The estimated integration orders for the 13 dynamic orthogonal components range from 0.0087 to 0.7328 and indicate that some of the components may have short memory while others behave like stationary or nonstationary fractionally integrated processes. We clearly reject equality of all integration orders, while also each of the groupings in two groups can be rejected on a 0.01 significance level. For three groups, we do not reject the hypothesis of equal integration orders within groups.
The sequential test for groups with equal memory yields $j^*=3$ with a p-value of 0.3181,
where groups of three ($\hat{d}^{(1)} = 0.6717$), seven ($\hat{d}^{(2)} = 0.3448$) and three ($\hat{d}^{(3)} = 0.0523$) components are identified, respectively. The hypothesis that $d^{(3)}=0$ is not rejected. We may therefore treat the members of the third group as short-range dependent and belonging to $\vz_t$. Thus, $s_1=3$, $s_2=7$ and $s_0=3$ appear as a reasonable specification for model \eqref{eq:DOC_FC} due to the preliminary analysis.

We obtain starting values for the parametric estimator from this procedure. Firstly, $\vd$ and $\vphi$ are estimated from the dynamic orthogonal components. Secondly, from regressing observed data on standardized estimated orthogonal components with unit innovation variance, we obtain starting values for $\vh$, $\mLambda$ and $\mGamma$, while certain columns of the latter matrices are rotated to satisfy the zero restrictions.

In very high-dimensional cases, the approach of \cite{PanYao2008} is not applicable, but \citet{LamYaoBa2011} and \citet{LamYao2012} provide feasible methods for stationary settings and comment on possible extensions to nonstationarity. In cases where the dynamic orthogonal components specification \eqref{eq:DOC_FC} is not appropriate, but the general setup \eqref{eq:FC} is, a specification search and preliminary estimates for the integration and cointegration parameters of the more general model could be based on the algorithm of \citet{Hua2009} which is capable of identifying and estimating cointegration subspaces by semiparametric methods.

\subsection{A parametric fractional components analysis} \label{sec:appl_fc}

We proceed with maximum likelihood estimation of the fractional components model using the EM algorithm of the state space representation. Although the exact state space respresentation is easily obtained using the current type II definition of fractional integration, the state dimension grows linearly with $n$ and becomes computationally infeasible. Instead, the latent fractionally integrated components are mapped to approximating ARMA(3,3) dynamics as described and justified by \citet{HarWei2018}. There, we show by simulation that low-order ARMA approximations (with parameters depending both on $d_j$ and on $n$) provide an excellent approximation performance and outperform truncated moving average and autoregressive representations by large amounts. 

We note that an asymptotic theory for maximum likelihood estimation in the fractionally cointegrated state space setup is not available. Certain functions of the parameter estimates are expected to exert nonstandard asymptotic behavior, especially in the nonstationary case $d_j>0.5$ for some $j$. However, normal and mixed normal asymptotics have been established and conventional tests and confidence intervals have been justified in different parametric fractional cointegration settings as well as in state space models with common unit root components \citep{ChaMilPa2009,ChaJiaPa2012}. We thus use standard parameter tests in what follows, bearing the preceding caveats in mind.

Constant terms are included by a further column $\vc$ in the observation matrix and estimated along with the free elements of $\mLambda$ and $\mGamma$. Setting the autoregressive order of $\vz_t$ to one and using starting values as described above, we estimate models with $q\in \{1,2,3\}$ groups of equal integration orders $d^{(j)}>0$ and additional autoregressive components. The Bayesian information criterion (BIC) is used to select sizes $s_0,\ldots,s_q$ and the value of $q$ with appropriate in-sample fit.\footnote{Instead of estimating all reasonable combinations of $s_0$, \ldots, $s_q$ for each $q$, we begin by the optimal grouping for a given $q$ obtained from the semiparametric methods of the previous section. From this specification, denoted as $s^{\{0\}}_j$, $j=0,\ldots,q$, we estimate all models characterized by $s_j \in \{s^{\{0\}}_j-1,s^{\{0\}}_j,s^{\{0\}}_j+1\}$, $j=0,\ldots,q$, given that they satisfy $s+s_0-1 \geq s_j^{\{0\}}\geq1$. The model with the least value of the BIC is selected and its indices denoted as $s^{\{1\}}_j$, and again models with indices close to $s^{\{1\}}_j$ are estimated and compared. This process is iterated until $s^{\{i\}}_j=s^{\{i-1\}}_j$ holds for all $j=0,\ldots,q$. As a result, also the number of white noise combinations may differ from 8, the result of the semiparametric analysis in the previous section.} We apply the BIC even if consistency is not established in this fractional setting. We expect that existing results hold for specification choices not involving the fractional components, while it is not clear to what extent the results of \citet{ChaJiaPa2012} carry over to the fractional setup. There, consistency of the BIC is shown for the number of stochastic trends in a unit root state space model.

We complement the semiparametric results of the previous section by a parametric specification search. After diagnostic checking of the selected model, we will take a closer look at its parameter estimates and implied long-run characteristics. The best models for each $q$ are shown in table \ref{tab:bic}, where estimated integration orders are given along with the log-likelihood (log-lik) and the BIC. Regarding the integration orders, we find that for $q>1$ estimates of $d^{(1)}$ are always above 0.5 suggesting nonstationarity of at least $s_1$ series in $\vy_t$. Overall, the models with $q=2$ are superior, in particular the grouping in $s_1=2$ and $s_2=9$ fractional and $s_0=2$ nonfractional components. This specification is similar to the one selected by the semiparametric approach and also suggests a dynamic dimension of $s+s_0=13$. Interestingly, the same specification with full noise covariance matrix $\mH$ is inferior ($BIC=-16.626$) as is the model with a full vector autoregressive matrix $\mPhi$ ($BIC=-17.150$). Furthermore, considering more lags in $\vz_t$ does not sufficiently improve the fit ($BIC=-17.155$ for $k=2$, $BIC=-17.046$ for $k=3$ and $BIC=-17.139$ for $k=4$).

\begin{figure}[h!]
\begin{center}\vspace{-0.5 cm}
\includegraphics[width=\textwidth]{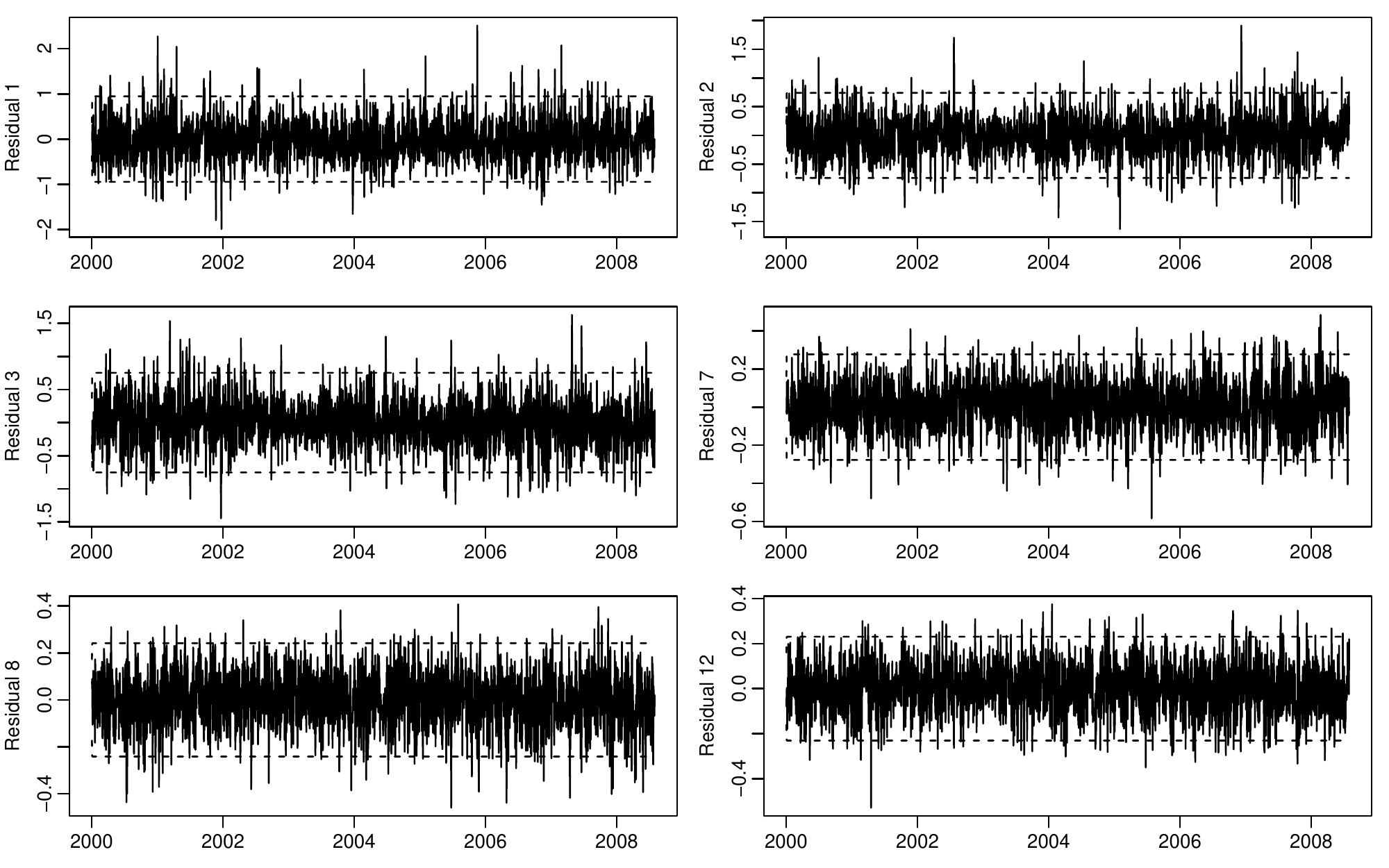}
\end{center}\vspace{-1cm} \caption{Residuals corresponding to log variances and z-transformed correlations for the first three assets for the fractional components model estimated in section \ref{sec:application}.}\label{fig:resid}
\end{figure}

\begin{figure}[h!]
\begin{center}\vspace{-0.5 cm}
\includegraphics[width=\textwidth]{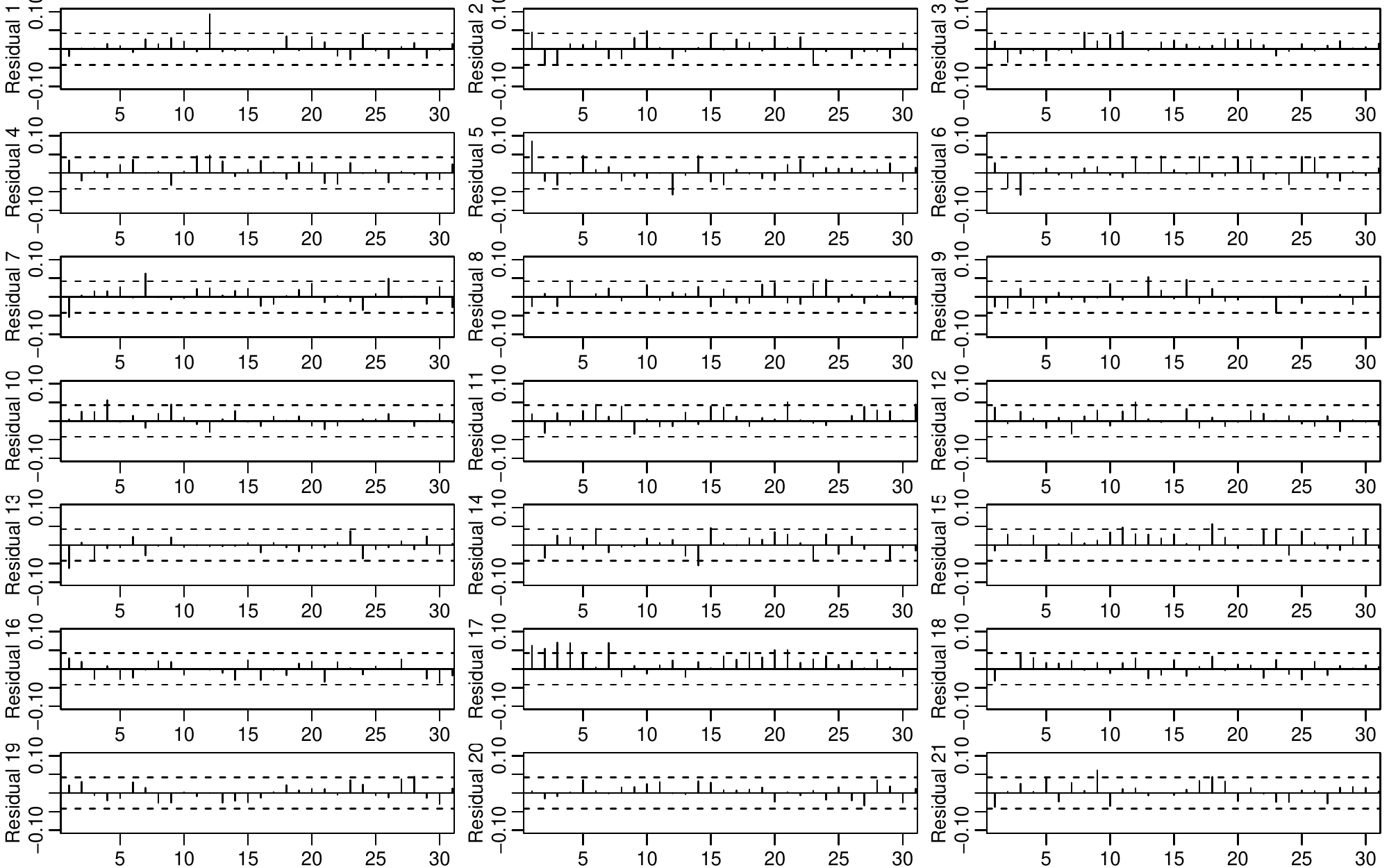}
\end{center}\vspace{-1cm} \caption{Residual autocorrelations for the fractional components model estimated in section \ref{sec:application}.}\label{fig:acf}
\end{figure}

\begin{figure}[h!]
\begin{center}\vspace{-0.5 cm}
\includegraphics[width=\textwidth]{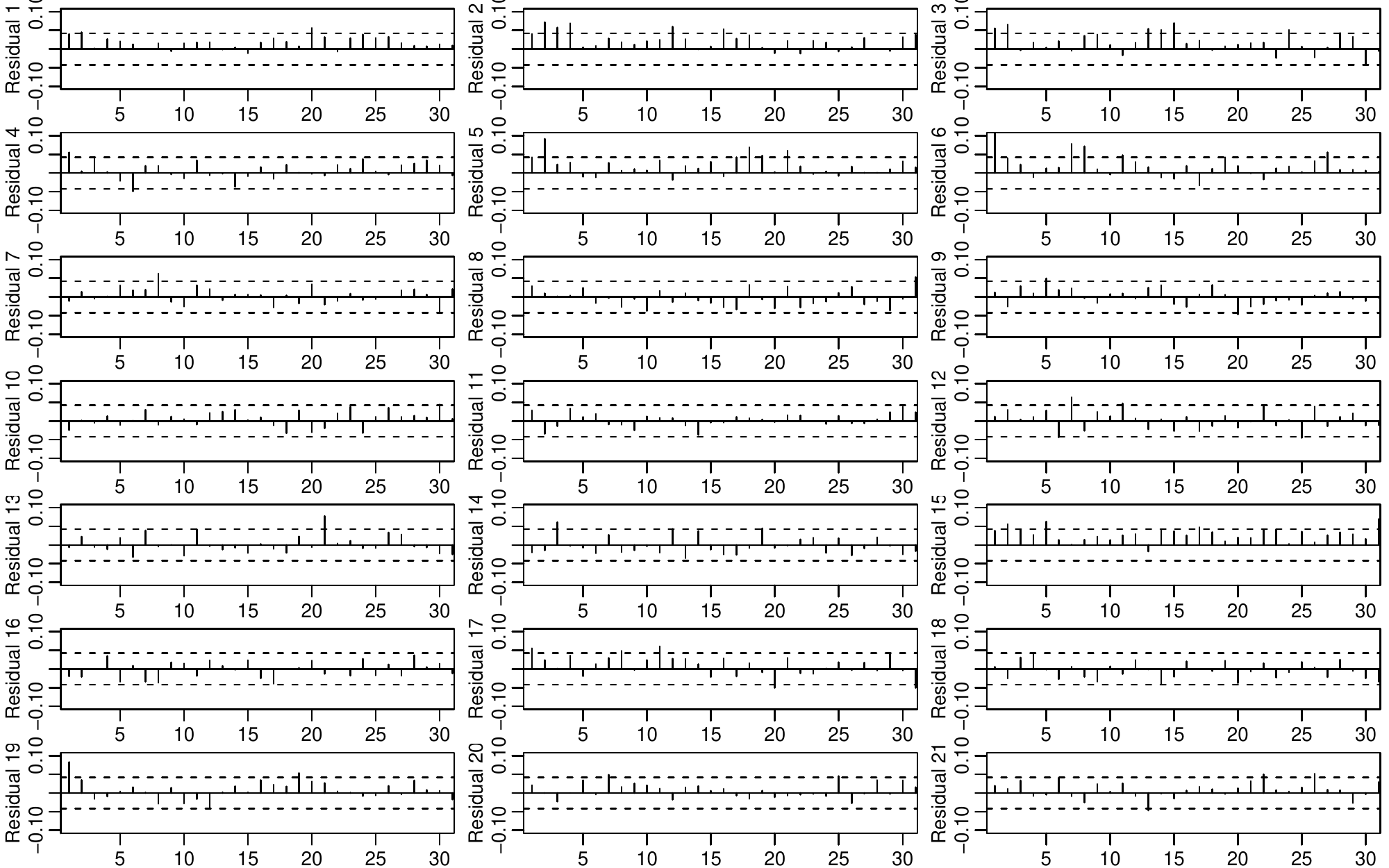}
\end{center}\vspace{-1cm} \caption{Autocorrelations of squared residuals for the fractional components model estimated in section \ref{sec:application}.}\label{fig:acf2}
\end{figure}

\begin{figure}[h!]
\begin{center}\vspace{-0.5 cm}
\includegraphics[width=\textwidth]{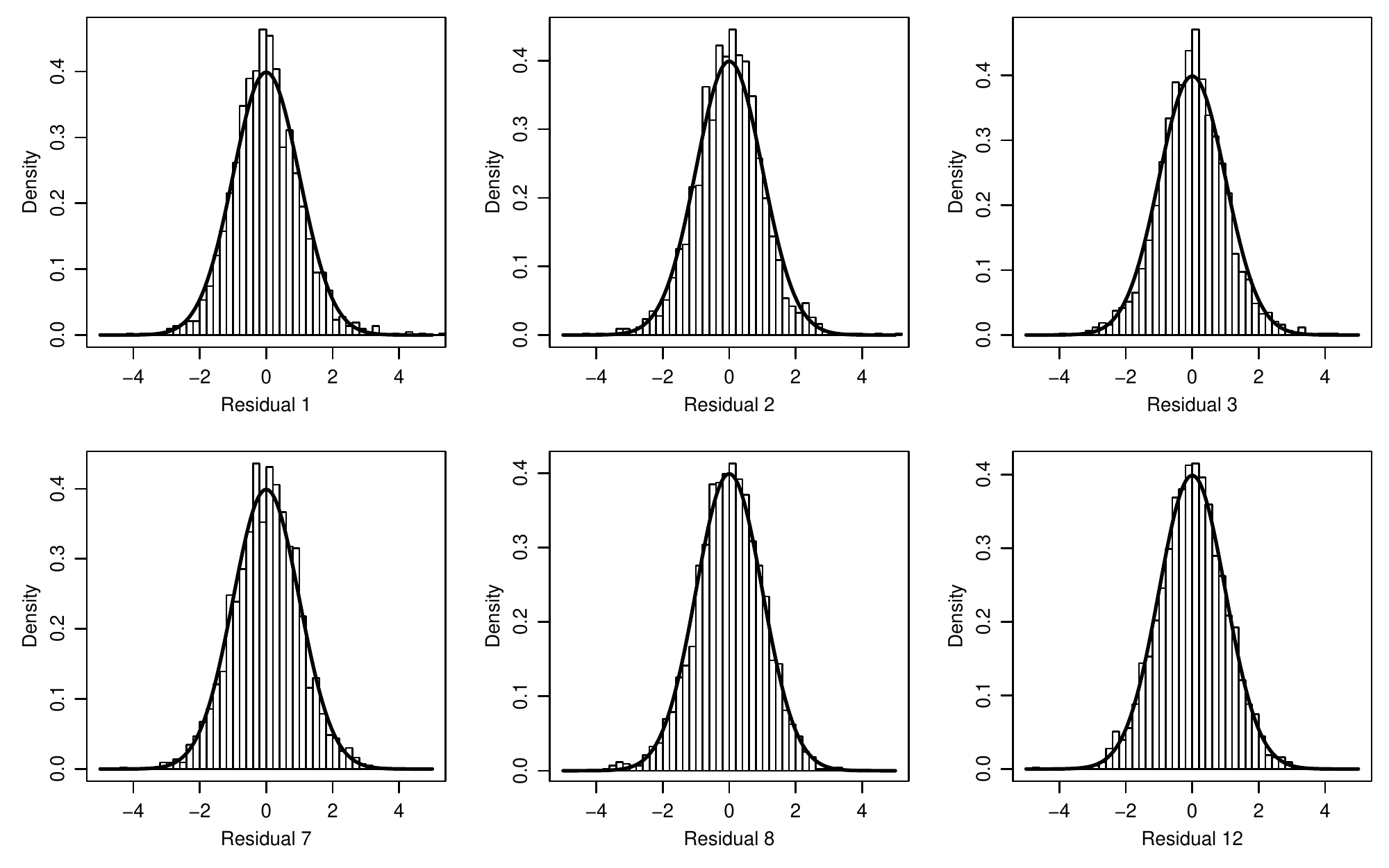}
\end{center}\vspace{-1cm} \caption{Histogram of residuals corresponding to log variances and z-transformed correlations for the first three assets for the fractional components model estimated in section \ref{sec:application} and normal density.}\label{fig:hist}
\end{figure}

We conduct several diagnostic tests on standardized model residuals $e_{it}= v_{it} / \sqrt{\mF_{ii,t}}$, where $\vv_t$ and $\mF_t$ are filtered residuals and forecast error covariance matrices, respectively. The residuals corresponding to log variances and z-transformed correlations for the first three assets are plotted in figure \ref{fig:resid}, while residual autocorrelations are depicted in figure \ref{fig:acf}, autocorrelations of squared residuals in figure \ref{fig:acf2} and histograms of the residuals along with the normal density in figure \ref{fig:hist}. The visual inspection shows some but no overwhelming evidence against the model assumptions. Autocorrelation both of residuals and squared residuals are generally below 0.1 in absolute value and mostly within the $\pm$ 2 standard error bands which are shown as horizontal lines. Some deviations from normality are visible, but not the sort of skewness and fat tails observed for models of untransformed residual variances and covariances.

Table \ref{tab:diagnostic} presents the diagnostic tests on standardized residuals. The p-values are shown for the Ljung-Box test (LM) and the ARCH-LM test for conditional heteroscedasticity (CH) for different lag length 5, 10 and 22. Additionally, the Jarque-Bera test result (JB) is shown in the last column. The null of no autocorrelation is not rejected at the 0.01 level for all but two or three residuals, depending on lag length. Clear evidence of conditional heteroskedasticity is found for the residuals of the log variance series, that is $e_{2t}$,$e_{3t}$, $e_{5t}$, and $e_{6t}$, where also the normality assumption is clearly rejected, but also for a few correlation series such as $e_{15,t}$ or $e_{19,t}$. A more flexible data transformation like the matrix Box-Cox approach of \citet{Wei2014} would typically ameliorate these findings, but we do not follow this approach further here.

Estimates of several of the model parameters are shown in table \ref{tab:estpar}. Along with the maximum likelihood estimates, we also show the mean of the estimators from a model-based bootstrap resampling exercise with 1000 iterations and generally find a low bias for the corresponding estimates. We also show standard errors, obtained in three ways, namely by the bootstrap (SE.boot), using the information matrix \citep[][section 3.4.5]{Har1991}, denoted by SE.info, and by the sandwich form \citet{Whi1982}, labelled SE.sand in the table. The different methods of computing standard errors give similar results, except for the variance parameters $h_i$, where the sandwich estimates are large compared to the others. Overall, including the parameters not shown in the table, the median ratio between bootstrap and sandwich standard errors is 1.31, while a typical sandwich estimate is 1.20 times larger than the corresponding estimate from the information matrix. We hence use the bootstrap methods in order to avoid a possible underestimation of the variances and spurious inference.

\begin{figure}[h!]
\begin{center}\vspace{-0.5 cm}
\includegraphics[width=\textwidth]{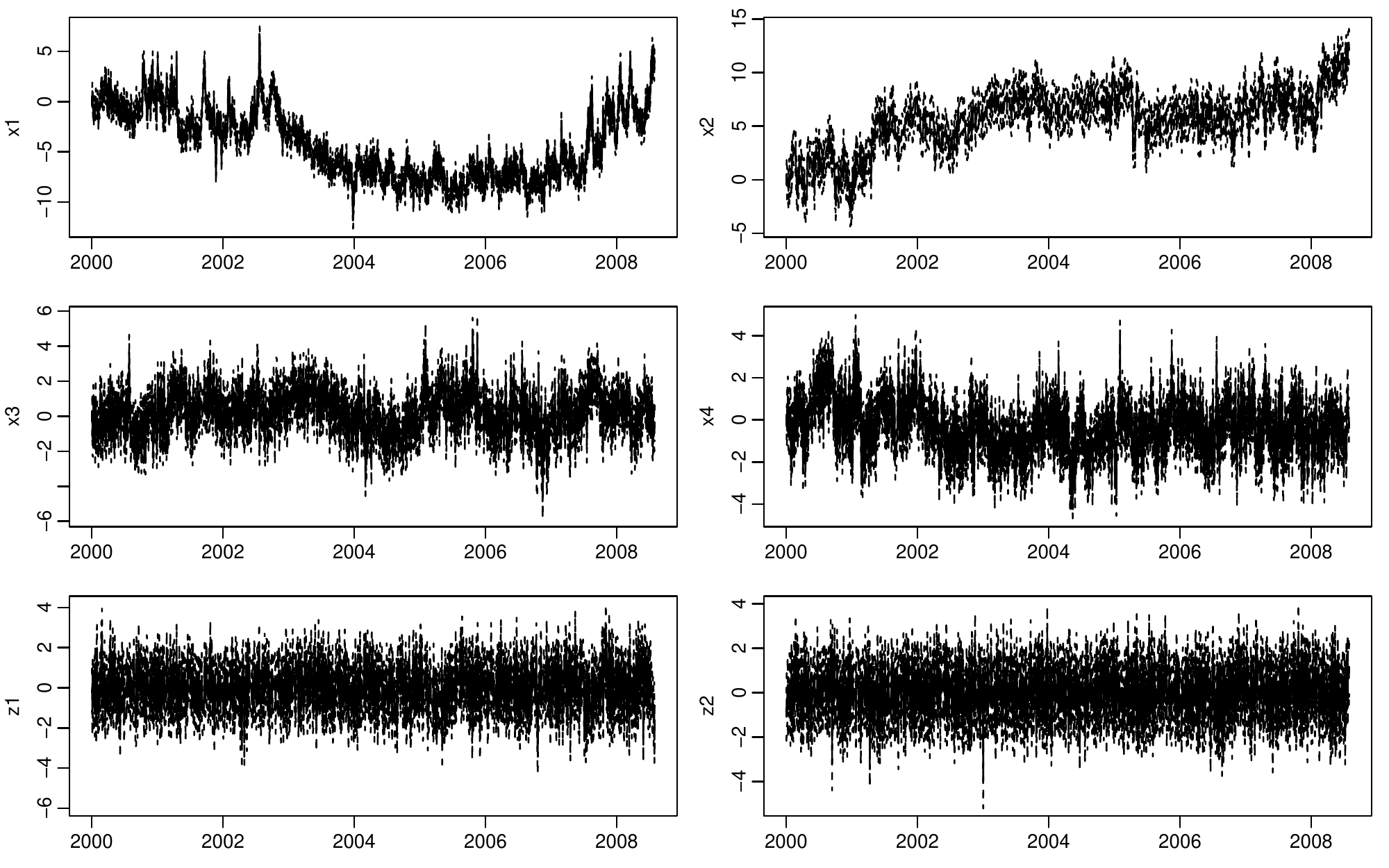}
\end{center}\vspace{-1cm} \caption{Selected smoothed fractional and nonfractional components (solid) $\pm$ 2 standard deviations (dashed) for the fractional components model estimated in section \ref{sec:application}. Both nonstationary components (above), the first two stationary long-memory components (middle) and the short-memory components (below) are given.}\label{fig:fac}
\end{figure}

The estimated memory parameters $d_1$ and $d_2$ exert a marked difference in the integration orders of fractional components. The two series in the first group are the cause of significant nonstationarity in our dataset. The second group of nine series introduces stationary long-memory persistence. In contrast, the nonfractional components in $\vz_t$ are only mildly autocorrelated, with small but significant autoregression parameters. Figure \ref{fig:fac} gives a visual impression of the factor dynamics, showing full sample (smoothed) estimates of the two nonstationary components (above), of the first two stationary long-memory components (middle) and of the short-memory components (below). The $\pm$ 2 standard error confidence intervals suggest a relatively precise estimation of the components. The different persistence of the three groups is clearly visible.

We turn to a discussion of the cointegration properties of the estimated system. In our preferred specification with a cointegration rank of $p-s_1=19$, and an 11-dimensional cointegration subspace, the loadings of fractional components provide an easier interpretation than the corresponding cointegration vectors, although the latter can be easily obtained and suitably normalized. 

With the abovementioned caveat that asymptotic properties are not available for this fractional cointegration setting, we show $t$-ratios for constants, for fractional loadings and for nonfractional loadings in table \ref{tab:tval_load}, where the bootstrap standard errors are used. The $t$-ratios for $\mLambda^{(1)}$ suggest that each of the series in $\vy_t$ is influenced by the nonstationary components, and hence all components of $\vy_t$ are nonstationary themselves. The first component loads very significantly on all variances with the same sign and can hence be interpreted as the main common risk factor. The second component represents joint common nonstationarity of the correlations, which is negatively associated with the IBM return variances. Except those corresponding to the first, the second and the forth stationary components with their equal signs, the columns of $\mLambda^{(2)}$ have a rather mixed pattern. Like the nonstationary factors, also the $I(d^{(2)})$ components affect variance and correlation dynamics at the same time and therefore induce fractional cointegration between log variances and z-transformed correlations. 

The finding of nonstationary fractional components affecting variances and correlations at the same time is new to the literature and may have remarkable consequences on portfolio selection and hedging opportunities, even at longer horizons. These effects should also be relevant to systemic risk measures as considered by central banks and regulators worldwide. To shed further light on the practical value of our approach, we turn to an evaluation of the forecasting precision in a real-world scenario in the following section.

\subsection{An out-of-sample comparison} \label{sec:oos}

We assess the forecasting performance of our model by means of an out-of-sample comparison. To avoid reference of the forecasts on the out-of-sample periods, we conduct a semiparametric specification search along the lines of section \ref{sec:semipar} for the first estimation sample only, i.e.\ for $\vy_t$, $t=1,\ldots,1508$, while $t=1509,\ldots,2156$ is reserved for prediction and therefore not used for selecting the specification. In this way, the model for the forecasting comparison includes $s_1=2$, $s_2=7$ and $s_0=3$ components of different integration orders. Rather than conducting comprehensive comparisons of a wide range of available methods which is beyond the scope of this paper, we select straightforward and simple benchmark models which have performed well in previous studies.

We choose the same out-of-sample setup as in \citet{Wei2014}. Thus, for each $T'\in [1508;2156-h]$, various competing models are estimated for a rolling sample with $n=1508$ observations, $\vy_{T'-1507}, \ldots, \vy_{T'}$. From these estimates, forecasts of $\vy_{T'+h}$, $h=1,5,10,20$, are computed. Also in line with \citet{Wei2014}, we compute bias-corrected forecasts of the realized covariance matrices $\hat{\mX}_{T'+h|T'}$ by the simulation-based technique discussed there. We evaluate the forecasting accuracy using the ex-post available data of the respective period.

The forecasting precision is assessed using different loss functions defined in appendix \ref{app:a3}. We consider the Frobenius norm $LF_{T',h}$ \eqref{eq:LF}, the Stein norm $LS_{T',h}$ \eqref{eq:LS} and the asymmetric loss $L3_{T',h}$ \eqref{eq:L3}; see \citet{LauRomVi2011} and \citet{LauRomVi2013}. Additionally, the ex-ante minimum variance portfolio is computed from the forecast and its realized variance $LMV_{T',h}$ \eqref{eq:LMV} used as a loss with obvious economic relevance. Furthermore, we assess density forecasts $f_r$ of the daily returns using covariance matrices, which are evaluated at the daily returns $r_{T'+h}$ in a logarithmic scoring rule $LD_{T',h}$ \eqref{eq:LD}.

As benchmarks, we consider two linear models for the log variance and z-transformed correlation series $\vy_t$, namely a diagonal vector ARMA(2,1) and a diagonal vector AR\-FI\-MA(1,$d$,1) model, which have been found to perform well by \citet{Wei2014}. Additionally, the diagonal vector AR\-FI\-MA(1,$d$,1) model is applied to the Cholesky factors of the covariance matrices \citep{ChiVoe2011}. Furthermore, we consider models with a conditional Wishart distribution, namely the conditional autoregressive Wishart (CAW) model of \citet{GolGriLi2012}, a dynamic correlation specification (CAW-DCC) of \citet{BauStoVi2012}, and additive and multiplicative components Wishart models as proposed by \citet{JinMah2013}. For further details on the comparison models consult appendix \ref{app:a3}.

For each loss function and horizon $h$, we compute the average losses (risks) for all models and obtain model confidence sets of \citet{HanLunNa2011}, bootstrapping the max-$t$ statistic with a block lengths of $\max\{5,h\}$. In tables \ref{tab:oos1}, \ref{tab:oos2}, \ref{tab:oos3} and \ref{tab:oos4}, we present the risks for $h=1,5,10,20$. The best performing model ($^{***}$) as well as members of the 80\% model confidence set ($^{**}$) and models contained in the 90\% but not in the 80\% set ($^{*}$) are indicated.

The fractional components model is among the best competitors for all horizons and loss functions. It has lowest risks for almost all setups. Exceptions occur for $h\geq10$ where the ARFIMA model for log variances and z-correlations performs best in some cases. Overall, the ARFIMA model on $\vy_t$ appears as second best in terms of forecasting precision.

The DOFC model is always contained in the 80\% model confidence set whereas all other models are rejected at least in some cases. For the Stein loss and the minimum-variance loss, the DOFC model is significantly superior than most competitors for small horizons, while with the Frobenius and asymmetric loss, rejections of other models are achieved for $h=10$ and $h=20$.

The performance of the fractional components model in terms of density forecasting is noteworthy. In each case there, our model is either the single member or one of two models in the confidence set and hence significantly outperforms most of the competitors. Since the behaviour of future daily returns is usually more important than the realized measures themselves, this finding is particularly strong from a practitioner's perspective.

Overall, we find a very good forecast performance of the model proposed in this paper. Although for some criteria and horizons statistical significance is lacking, the model yields very precise forecasts in relation to different competitors for all considered horizons and for several ways to measure this precision.

\section{Conclusion} \label{sec:conclusion}

We have suggested a general setup and a parsimonious model with very general fractional integration and cointegration properties. We discussed the usefulness of our approach for multivariate realized volatility modeling. In our application it was shown to provide a reasonable in-sample fit and competitive out-of-sample forecasting accuracy.

Several questions remain for further research. From an empirical point of view, we have shown the relevance of a very restricted specification in financial econometrics, but the general setup we introduced has a broader scope. Fractional components models with rich short-run dynamics may be considered for models of smaller dimension. In several empirical setups, fractional integration and cointegration has been found relevant, so that dynamic modeling, forecasting, identification of structural shocks and impulse response analyses in an according framework is a fruitful direction of ongoing research.

\section*{Acknowledgements}

The research of Roland Weigand has mostly been done at the Institute of Economics and Econometrics of the University of Regensburg and at the Institute for Labour Market Research (IAB) in Nuremberg. Very valuable comments by Rolf Tschernig, by Enzo Weber and by participants of the Interdisciplinary Workshop on Multivariate Time Series Modeling 2011 in Louvain La Neuve, at the Statistische Woche 2011 in Leipzig, and of research seminars at the Universities of Regensburg, Augsburg and Bielefeld are gratefully acknowledged. The authors are also thankful to Niels Aka for providing R codes to estimate model confidence sets. Tobias Hartl gratefully acknowledges support through the projects TS283/1-1 and WE4847/4-1 financed by the German Research Foundation (DFG).

\newpage
\begin{spacing}{1.2}
\bibliographystyle{dcu}
\bibliography{../literatur/literatur}
\end{spacing}
\newpage
\appendix

\section{Details on alternative representations}\label{app:a1}

In this appendix we provide more details on the derivation of the alternative representations of the fractional components model \eqref{eq:FC} which we discuss in section \ref{sec:alternative}.

To derive the error correction representation \eqref{eq:vecm}, we start from the FC setup with $q=2$ and $s=p$,
\[
\vy_t = \mLambda^{(1)} \vx_t^{(1)} + \mLambda^{(2)} \vx_t^{(2)} + \vu_t,
\]
from which we note that
\[
\mLambda_{\bot}^{(1)'} \Delta^{d^{(2)}} \vy_t = \mLambda_{\bot}^{(1)'} \mLambda^{(2)} \vxi_t^{(2)} + \mLambda_{\bot}^{(1)'} \Delta^{d^{(2)}}\vu_t\quad \text{and} \quad \mLambda_{\bot}^{(2)'} \Delta^{d^{(1)}} \vy_t = \mLambda_{\bot}^{(2)'} \mLambda^{(1)} \vxi_t^{(1)} + \mLambda_{\bot}^{(2)'} \Delta^{d^{(1)}}\vu_t.
\]
We define
\beq \label{eq:defm}
\mN:=\mLambda^{(2)}(\mLambda^{(1)'}_{\bot} \mLambda^{(2)})^{-1}\mLambda^{(1)'}_{\bot} \quad \text{and} \quad \mM:=\mLambda^{(1)}(\mLambda^{(2)'}_{\bot} \mLambda^{(1)})^{-1}\mLambda^{(2)'}_{\bot},
\eeq
and make use of $\mI =\mN+\mM$ \citep{Joh2008}, to obtain
\beq \label{eq:vecm1}
\Delta^{d^{(1)}} \vy_t =  \mM(\mLambda^{(1)}\vxi_t^{(1)} + \Delta^{d^{(1)}} \vu_t ) + \Delta^{d^{(1)}-d^{(2)}}\Delta^{d^{(2)}} \mN\vy_t.
\eeq
Adding and substracting $\Delta^{d^{(2)}} \mN\vy_t$ on the right side of \eqref{eq:vecm1} and the decomposition $\mN=-\valpha\beta'$ yields \eqref{eq:vecm}.


Next, we consider the triangular representation; see \eqref{eq:tria1} and \eqref{eq:triaj}. The first block, \eqref{eq:tria1}, is easily obtained. Since $\mLambda^{(1,1)}$ is nonsingular and we also assumed a nonsingular covariance matrix of the white noise sequence $\vxi_t$, we find that the first term on the right is $I(0)$ with positive definite spectral density while the other terms have integration orders lower than zero, leading to $\vomega_t^{(1)} \sim I(0)$.
To arrive at the $j$-th block of the system, consider the expression for $\vy_t^{(j)}$,
\[
\Delta^{d^{(j)}}\vy_t^{(j)} = \mLambda^{(j,1)} \Delta^{d^{(j)}}\vx_t^{(1)} + \ldots +\mLambda^{(j,q)} \Delta^{d^{(j)}}\vx_t^{(q)} + \Delta^{d^{(j)}}\vu_t^{(j)}.
\]
Since $\Delta^{d^{(j)}}\vx_t^{(i)}$ is integrated of order zero or lower for $i\geq j$, we can write
\begin{align} \label{eq:tria2}
\Delta^{d^{(j)}}\vy_t^{(j)} &= \mLambda^{(j,1)} \Delta^{d^{(j)}}\vx_t^{(1)} + \ldots +\mLambda^{(j,j-1)} \Delta^{d^{(j)}}\vx_t^{(j-1)} + \tilde{\vomega}_t^{j} \nonumber\\
&= \mLambda^{(j,1:(j-1))} \Delta^{d^{(j)}}\vx_t^{(1:(j-1))} + \tilde{\vomega}_t^{j},
\end{align}
where $\tilde{\vomega}_t^{j} \sim I(0)$.
To substitute for the latent variables in this expression, consider
\[
\Delta^{d^{(j)}}\vy_t^{(1:(j-1))} = \mLambda^{(1:(j-1),1:(j-1))} \Delta^{d^{(j)}}\vx_t^{(1:(j-1))} + \check{\vomega}_t^{j},
\]
with $\check{\vomega}_t^{j} \sim \; I(0)$ which we can solve for
\[
\Delta^{d^{(j)}}\vx_t^{(1:(j-1))} = (\mLambda^{(1:(j-1),1:(j-1))})^{-1} \Delta^{d^{(j)}}\vy_t^{(1:(j-1))} - (\mLambda^{(1:(j-1),1:(j-1))})^{-1} \check{\vomega}_t^{j}.
\]
Substituting this expression into \eqref{eq:tria2} yields the general expression \eqref{eq:triaj} for the $j$-th block of the triangular system for $j=2,\ldots,q$, where
\[
\vomega_t^{(j)} = \tilde{\vomega}_t^{j} - \mLambda^{(j,1:(j-1))} (\mLambda^{(1:(j-1),1:(j-1))})^{-1} \check{\vomega}_t^{j},
\]
which can be stated in greater detail as
\begin{multline} \label{eq:omega}
\vomega_t^{(j)} = -\mLambda^{(j,1:(j-1))} (\mLambda^{(1:(j-1),1:(j-1))})^{-1} \mLambda^{(1:(j-1),j:q)} \Delta^{d^{(j)}}\vx_t^{(j:q)} \\ - \mLambda^{(j,1:(j-1))} (\mLambda^{(1:(j-1),1:(j-1))})^{-1} \Delta^{d^{(j)}}\vu_t^{(1:(j-1))} + \mLambda^{(j,j:q)} \Delta^{d^{(j)}}\vx_t^{(j:q)} + \Delta^{d^{(j)}}\vu_t^{(j)}.
\end{multline}
This process is the sum of several additive negatively integrated plus a white noise process
\[
\left[\mLambda^{(j,j)}  - \mLambda^{(j,1:(j-1))} (\mLambda^{(1:(j-1),1:(j-1))})^{-1} \mLambda^{(1:(j-1),j)} \right] \vxi_t^{(j)},
\]
so that we conclude that $\vomega_t^{(j)}$ is $I(0)$ with positive definite spectral density at zero frequency.

We arrive at the representation \eqref{eq:triaB} where $\mB$ is partitioned into blocks according to
\[
\mB =
\bmat
\mI & \mzeros & \ldots & \mzeros \\
\mB^{(1,1)} & \mI &   & \mzeros \\
\vdots & \ddots & \ddots & \vdots \\
\mB^{(q,1)} & \ldots & \mB^{(q,q-1)} & \mI
\emat.
\]

In case $p>s$, we have
\begin{align*}
\vy_t^{(q+1)} &= \mLambda^{(q+1,1:q)} (\mLambda^{(1:q,1:q)})^{-1} \vy_t^{(1:q)}  + \vu_t^{(q+1)} -  \mLambda^{(q+1,1:q)} (\mLambda^{(1:j,1:j)})^{-1} \vu_t^{(1:q)}\\
&= \mB^{(q+1,1)} \vy^{(1)}_t + \ldots + \mB^{(q+1,q)} \vy^{(j-1)}_t + \vomega_t^{(q+1)},
\end{align*}
and the representation \eqref{eq:triaB} is changed to
\[
\mB \vy_t = (\Delta_+^{-d_1}\vomega_t^{(1)}, \ldots,\Delta_+^{-d_q}\vomega_t^{(q)}, \vomega_t^{(q+1)})'
\]
where $\mB$ is extended by the $p-s$ rows $(\mB^{(q+1,1)},\ldots, \mB^{(q+1,q)}, \mI)$.

\section{Details on the out-of-sample comparison} \label{app:a3}

In this section we give further details on the out-of-sample evaluation of section \ref{sec:oos}. We state the loss functions to evaluate the forecasts as well as the specifications of the benchmark models and their estimation.

For given forecasted realized covariance matrices $X_{T'+h|T'}$ and realizations $X_{T'+h}$, the loss functions considered in this paper are the Frobenius norm ($LF_{T',h} $), the Stein distance ($LS_{T',h} $), the asymmetric loss ($L3_{T',h} $), the realized variance of the ex-ante minimum variance portfolio ($LMV_{T',h} $), and the negative log-score of density forecasts $f_r$ ($LD_{T',h} $), given by
\begin{align}
LF_{T',h} &= {\textstyle \sum_{i=1}^k\sum_{j=1}^k}(\mX_{ij,T'+h}-\mX_{ij,T'+h|T'})^2, \label{eq:LF}\\
LS_{T',h} &= \tr\left[X_{T'+h|T'}^{-1} X_{T'+h} \right]- \log\left|X_{T'+h|T'}^{-1} X_{T'+h} \right|-k, \label{eq:LS} \\
L3_{T',h} &= \frac{1}{6}\tr\left[X_{T'+h|T'}^{3} - X_{T'+h}^{3} \right]- \frac{1}{2} \tr \left[ X_{T'+h|T'}^{2} (X_{T'+h}-X_{T'+h|T'}) \right], \label{eq:L3}\\
LMV_{T',h} &= w'X_{T'+h}w, \quad w=(\iota' X_{T'+h|T'} \iota)^{-1} X_{T'+h|T'}\iota, \quad \iota=(1,\ldots,1)', \label{eq:LMV}\\
LD_{T',h} &= - \log f_r(r_{T'+h}). \label{eq:LD}
\end{align}

As comparison models we consider three linear models in transformed covariance matrices, namely the diagonal vector ARMA(2,1) model
\beq \label{eq:ARMA}
(1- \phi_{i1}L-\phi_{i2}L^2) (y_{it}-c_i)=(1+\theta_{i1}L) v_{it}, \quad i=1,\ldots,21,
\eeq
for the log variance and z-correlation series $\vy_t$, a diagonal vector ARFIMA(1,$d$,1) model
\beq \label{eq:ARFIMA}
(1- \phi_{i1}L)(1-L)^{d_i} (y_{it}-c_i)=(1+\theta_{i1}L) v_{it}, \quad i=1,\ldots,21,
\eeq
for $\vy_t$ and the same model \eqref{eq:ARFIMA} applied to Cholesky factors. The same model orders have been used by \citet{ChiVoe2011} and \citet{Wei2014} and were found to compete favorably with other choices. The dynamic parameters of these models are estimated by Gaussian quasi maximum likelihood equation by equation, with no cross-equation restrictions such as equality of memory parameters. A full covariance matrix of the error terms is estimated from the residuals.

The other four benchmark models are based on a conditional Wishart distribution,
\begin{equation} \label{eq:wish}
X_t | {\cal I}_{t-1} \sim W_n(\nu, S_t/\nu),
\end{equation}
where $\calI_t$ is the information set consisting of $\mX_s$, $s\leq t$, $W_n$ denotes the central Wishart density, $\nu$ is the scalar degrees of freedom parameter and $S_t/\nu$ is a $(6 \times 6)$ positive definite scale matrix, which is related to the conditional mean of $X_t$ by $E[X_t|{\cal I}_{t-1}]=S_t$.
The baseline CAW(p,q) model of \citet{GolGriLi2012} specifies the conditional mean as
\beq \label{eq:CAW}
S_t = CC' + \sum_{j=1}^p B_j S_{t-j} B_j' + \sum_{j=1}^q A_j X_{t-j} A_j',
\eeq
$C$, $B_j$ and $A_j$ denoting $(6 \times 6)$ parameter matrices, while the CAW-DCC model of \citet{BauStoVi2012} employs a decomposition $S_t = H_tP_tH_t'$ where $H_t$ is diagonal and $P_t$ is a well-defined correlation matrix. As a sparse and simple DCC benchmark we apply univariate realized GARCH($p_v$,$q_v$) specifications for the realized variances
\beq \label{eq:CAW-DCC2}
H_{ii,t}^2 = c_i + \sum_{j=1}^{p_v} b^v_{i,j} H_{ii,t-j}^2 + \sum_{j=1}^{q_v} a^v_{i,j} X_{ii,t-j},
\eeq
along with the `scalar Re-DCC' model \citep{BauStoVi2012} for the realized correlation matrix $R_t$,
\beq \label{eq:CAW-DCC3}
P_t = \bar{P} + \sum_{j=1}^{p_c} b^c_{j} P_{t-j} + \sum_{j=1}^{q_c} a^c_{j} R_{t-j}.
\eeq
The diagonal CAW($p$,$q$) and the CAW-DCC($p$,$q$) specification with $p=p_v=p_c=2$ and $q=q_v=q_c=1$ are selected since they provide a reasonable in-sample fit among various order choices. They are estimated by maximum likelihood using variance and correlation targeting.

\clearpage
\renewcommand{\arraystretch}{1.3}
%
\clearpage

\begin{table}[ht]
\centering
\begin{tabular}{rrrr|rrrrr}
$s_1$ & $s_2$ & $s_3$ & $s_0$ & log-lik & $d^{(1)}$ & $d^{(2)}$ & $d^{(3)}$ & BIC \\
  \hline
12 &  &  & 3 & -19572.2 & 0.368 &  &  & -17.116 \\
  10 &  &  & 4 & -19553.8 & 0.390 &  &  & -17.106 \\
  11 &  &  & 4 & -19590.6 & 0.390 &  &  & -17.101 \\
  12 &  &  & 4 & -19620.7 & 0.383 &  &  & -17.094 \\
  10 &  &  & 5 & -19601.8 & 0.411 &  &  & -17.087 \\
  11 &  &  & 5 & -19636.4 & 0.406 &  &  & -17.080 \\
  2 & 9 &  & 2 & -19573.8 & 0.631 & 0.338 &  & -17.157 \\
  2 & 10 &  & 1 & -19538.0 & 0.596 & 0.319 &  & -17.156 \\
  1 & 10 &  & 2 & -19530.9 & 0.653 & 0.370 &  & -17.146 \\
  2 & 10 &  & 2 & -19605.0 & 0.551 & 0.303 &  & -17.143 \\
  3 & 9 &  & 1 & -19546.7 & 0.619 & 0.315 &  & -17.139 \\
  2 & 8 &  & 3 & -19575.9 & 0.634 & 0.353 &  & -17.134 \\
  2 & 2 & 7 & 2 & -19600.7 & 0.639 & 0.422 & 0.304 & -17.129 \\
  2 & 3 & 7 & 2 & -19666.6 & 0.565 & 0.417 & 0.252 & -17.122 \\
  2 & 3 & 6 & 2 & -19607.5 & 0.634 & 0.407 & 0.288 & -17.121 \\
  2 & 4 & 6 & 2 & -19676.5 & 0.634 & 0.412 & 0.234 & -17.121 \\
  3 & 3 & 5 & 2 & -19617.4 & 0.629 & 0.398 & 0.272 & -17.119 \\
  3 & 2 & 6 & 2 & -19601.0 & 0.618 & 0.395 & 0.292 & -17.115 \\
\end{tabular}\caption{Estimation results for different specifications of the models estimated in section \ref{sec:appl_fc}. We show the combinations of $s_j$, $j=0,\ldots,q$ with best values of the BIC for $q=1$ (above), $q=2$ (middle) and $q=3$ (below).} \label{tab:bic}
\end{table}

\begin{table}[ht]
\centering
\begin{tabular}{l|rrrrrrr}
 & LB5 & LB10 & LB22 & CH5 & CH10 & CH22 & JB \\
  \hline
  $e_{1,t}$ & 0.944 & 0.848 & 0.101 & 0.083 & 0.359 & 0.373 & 0.000 \\
  $e_{2,t}$ & 0.022 & 0.008 & 0.032 & 0.000 & 0.001 & 0.001 & 0.000 \\
  $e_{3,t}$ & 0.191 & 0.110 & 0.253 & 0.008 & 0.018 & 0.006 & 0.000 \\
  $e_{4,t}$ & 0.474 & 0.459 & 0.109 & 0.043 & 0.038 & 0.152 & 0.000 \\
  $e_{5,t}$ & 0.000 & 0.004 & 0.002 & 0.000 & 0.004 & 0.000 & 0.000 \\
  $e_{6,t}$ & 0.035 & 0.197 & 0.063 & 0.000 & 0.000 & 0.000 & 0.000 \\ 
  $e_{7,t}$ & 0.091 & 0.054 & 0.178 & 0.741 & 0.142 & 0.382 & 0.002 \\
  $e_{8,t}$ & 0.071 & 0.075 & 0.103 & 0.587 & 0.569 & 0.509 & 0.000 \\
  $e_{9,t}$ & 0.208 & 0.365 & 0.295 & 0.109 & 0.280 & 0.212 & 0.219 \\
  $e_{10,t}$ & 0.108 & 0.117 & 0.459 & 0.861 & 0.915 & 0.717 & 0.001 \\
  $e_{11,t}$ & 0.326 & 0.090 & 0.092 & 0.207 & 0.436 & 0.877 & 0.000 \\
  $e_{12,t}$ & 0.468 & 0.477 & 0.442 & 0.538 & 0.033 & 0.037 & 0.175 \\
  $e_{13,t}$ & 0.080 & 0.158 & 0.800 & 0.571 & 0.318 & 0.060 & 0.000 \\
  $e_{14,t}$ & 0.235 & 0.162 & 0.026 & 0.080 & 0.167 & 0.079 & 0.000 \\
  $e_{15,t}$ & 0.242 & 0.328 & 0.072 & 0.001 & 0.011 & 0.011 & 0.102 \\
  $e_{16,t}$ & 0.354 & 0.541 & 0.589 & 0.272 & 0.180 & 0.367 & 0.000 \\
  $e_{17,t}$ & 0.000 & 0.000 & 0.000 & 0.057 & 0.039 & 0.003 & 0.369 \\
  $e_{18,t}$ & 0.158 & 0.376 & 0.480 & 0.245 & 0.326 & 0.349 & 0.000 \\
  $e_{19,t}$ & 0.557 & 0.514 & 0.849 & 0.003 & 0.011 & 0.019 & 0.001 \\
  $e_{20,t}$ & 0.685 & 0.882 & 0.942 & 0.412 & 0.216 & 0.790 & 0.000 \\
  $e_{21,t}$ & 0.122 & 0.014 & 0.055 & 0.600 & 0.446 & 0.256 & 0.000 \\
\end{tabular} \caption{P-values of diagnostic tests for the residuals from the DOFC model \eqref{eq:DOC_FC} estimated in section \ref{sec:appl_fc}. We conducted Ljung-Box tests for residual correlation (LB), ARCH-LM tests for conditional heteroskedasticity (CH), each with different lags, and Jarque-Bera tests (JB) for deviations from normality.} \label{tab:diagnostic}
\end{table}

\begin{table}[ht]
\centering
\begin{tabular}{l|rrrrr}
& Estimate & Mean & SE.boot & SE.sand & SE.info \\
  \hline
  $d_1$ & 0.6308 & 0.6361 & 0.0190 & 0.0217 & 0.0178 \\
  $d_2$ & 0.3382 & 0.3334 & 0.0094 & 0.0116 & 0.0086 \\
  $\phi_1$ & 0.2468 & 0.2360 & 0.0345 & 0.0417 & 0.0348 \\
  $\phi_2$ & 0.0768 & 0.0636 & 0.0370 & 0.0419 & 0.0402 \\    
  $h_{1}$& 0.2028 & 0.1844 & 0.1122 & 0.1106 & 0.0759 \\
  $h_{2}$& 0.3858 & 0.3727 & 0.0522 & 0.0551 & 0.0321 \\
  $h_{3}$& 0.3289 & 0.3309 & 0.0930 & 0.0957 & 0.0714 \\
  $h_{4}$& 0.1758 & 0.1638 & 0.1222 & 0.1371 & 0.0861 \\
  $h_{5}$& 0.7649 & 0.7618 & 0.0558 & 0.0676 & 0.0482 \\
  $h_{6}$& 0.2459 & 0.2413 & 0.0772 & 0.0810 & 0.0588 \\    
  $h_{7}$& 0.0615 & 0.0611 & 0.0037 & 0.0079 & 0.0027 \\
  $h_{8}$& 0.0746 & 0.0739 & 0.0032 & 0.0063 & 0.0026 \\
  $h_{9}$& 0.0799 & 0.0793 & 0.0033 & 0.0060 & 0.0027 \\
  $h_{10}$ & 0.0778 & 0.0771 & 0.0034 & 0.0060 & 0.0028 \\
  $h_{11}$ & 0.0725 & 0.0718 & 0.0036 & 0.0072 & 0.0030 \\
  $h_{12}$ & 0.0563 & 0.0557 & 0.0036 & 0.0062 & 0.0025 \\
  $h_{13}$ & 0.0545 & 0.0543 & 0.0032 & 0.0063 & 0.0026 \\
  $h_{14}$ & 0.0509 & 0.0505 & 0.0031 & 0.0060 & 0.0025 \\
  $h_{15}$ & 0.0570 & 0.0564 & 0.0056 & 0.0077 & 0.0045 \\
  $h_{16}$ & 0.0739 & 0.0733 & 0.0033 & 0.0059 & 0.0029 \\
  $h_{17}$ & 0.0889 & 0.0880 & 0.0036 & 0.0053 & 0.0032 \\
  $h_{18}$ & 0.0441 & 0.0438 & 0.0040 & 0.0082 & 0.0030 \\
  $h_{19}$ & 0.0919 & 0.0910 & 0.0038 & 0.0059 & 0.0035 \\
  $h_{20}$ & 0.0621 & 0.0615 & 0.0037 & 0.0064 & 0.0031 \\
  $h_{21}$ & 0.0601 & 0.0595 & 0.0035 & 0.0060 & 0.0032 \\
\end{tabular}\caption{Estimated parameters along with bootstrap mean and standard errors from bootstrap (SE.boot), sandwich (SE.sand) and information matrix (SE.info) as described in \cite{HarWei2018} for the DOFC model \eqref{eq:DOC_FC} estimated in section \ref{sec:appl_fc}.} \label{tab:estpar}
\end{table}

\begin{table}[ht]
\centering
\begin{tabular}{l|rr|rrrrrrrrr|rr}
& \multicolumn{2}{c|}{$\mLambda^{(1)}$} & \multicolumn{9}{c|}{$\mLambda^{(2)}$} & \multicolumn{2}{c}{$\mGamma$} \\
  \hline
 $y_{1 ,t}$ & 17.4 &  & 11.9 &  &  &  &  &  &  &  &  & 3.6 &  \\
 $y_{2 ,t}$ & 22.7 & 1.6 & 1.8 & -11.9 &  &  &  &  &  &  &  & 0.4 & -0.7 \\
 $y_{3 ,t}$ & 13.1 & -1.6 & 3.4 & -6.1 & -15.6 &  &  &  &  &  &  & 1.1 & 1.3 \\
 $y_{4 ,t}$ & 11.9 & -2.9 & 2.1 & -3.3 & -5.2 & 16.2 &  &  &  &  &  & 0.6 & -1.6 \\
 $y_{5 ,t}$ & 12.5 & -11.1 & 4.4 & -10.7 & -1.1 & 1.1 & -3.4 &  &  &  &  & 1.6 & -1.7 \\
 $y_{6 ,t}$ & 18.9 & 1.2 & 2.0 & -6.0 & -2.3 & 3.1 & -3.1 & 6.8 &  &  &  & 1.0 & 1.7 \\   
 $y_{7 ,t}$ & 4.1 & 6.0 & 5.6 & -8.0 & -1.1 & 2.0 & -4.7 & -1.5 & 4.9 &  &  & -0.5 & -2.7 \\
 $y_{8 ,t}$ & 3.5 & 3.2 & 5.4 & -10.2 & -1.2 & 2.4 & -1.6 & 0.5 & 4.4 & 7.4 &  & 0.9 & 0.2 \\
 $y_{9 ,t}$ & 4.7 & 4.2 & 2.8 & -7.4 & -1.0 & 2.2 & -5.8 & -2.2 & -0.1 & 3.4 & -3.0 & 3.9 & -0.5 \\
 $y_{10,t}$ & 2.7 & 4.9 & 2.5 & -9.0 & -2.9 & 2.0 & -3.3 & -0.9 & 2.5 & 3.6 & 4.2 & 6.1 & -4.0 \\
 $y_{11,t}$ & 4.0 & 5.1 & 3.5 & -9.9 & 0.6 & 2.7 & -2.8 & -1.7 & 7.1 & -1.8 & -2.5 & 3.2 & 1.7 \\
 $y_{12,t}$ & 2.9 & 3.8 & 9.7 & -9.8 & -1.0 & 2.5 & -1.3 & 0.2 & 0.5 & 3.5 & 3.5 & -2.5 & 3.0 \\
 $y_{13,t}$ & 3.9 & 4.3 & 5.5 & -7.0 & -0.9 & 2.2 & -5.7 & -2.7 & -3.1 & 0.6 & -1.0 & 0.2 & 2.7 \\
 $y_{14,t}$ & 2.0 & 4.4 & 5.1 & -9.8 & -2.6 & 2.6 & -2.3 & -0.8 & -1.8 & -0.8 & 6.1 & 2.6 & -3.4 \\
 $y_{15,t}$ & 3.4 & 6.0 & 6.2 & -12.5 & 0.5 & 2.5 & -1.6 & 0.6 & 1.4 & -7.2 & -0.4 & -0.3 & 3.5 \\
 $y_{16,t}$ & 4.4 & 3.1 & 5.6 & -10.6 & -1.1 & 2.3 & -1.7 & -0.4 & -2.9 & 5.4 & -2.2 & 1.2 & 2.2 \\
 $y_{17,t}$ & 3.2 & 4.4 & 5.7 & -12.5 & -1.5 & 2.8 & 0.5 & 1.7 & -0.8 & 3.9 & 4.2 & 3.8 & -0.9 \\
 $y_{18,t}$ & 2.5 & 3.1 & 6.6 & -12.6 & 0.0 & 2.8 & 0.4 & 0.3 & 2.3 & 2.2 & 1.2 & 1.5 & 10.2 \\
 $y_{19,t}$ & 3.7 & 5.2 & 3.6 & -8.9 & -1.9 & 1.7 & -2.6 & -0.9 & -5.1 & 2.1 & -0.7 & 6.2 & -1.9 \\
 $y_{20,t}$ & 3.7 & 4.0 & 3.4 & -8.1 & 0.1 & 2.6 & -4.7 & -3.2 & -2.4 & 0.2 & -2.2 & 3.9 & 6.4 \\
 $y_{21,t}$ & 1.6 & 4.0 & 3.1 & -9.9 & -1.4 & 3.0 & -1.2 & -0.5 & -0.3 & -2.1 & 4.5 & 8.0 & 1.4 \\
\end{tabular} \caption{Bootstrap $t$-ratios for fractional components loadings ($\mLambda^{(1)}$ and $\mLambda^{(2)}$) and nonfractional loadings ($\mGamma$) from the DOFC model \eqref{eq:DOC_FC} estimated in section \ref{sec:appl_fc}.} \label{tab:tval_load}
\end{table}

\begin{table}[ht]
\centering
\begin{tabular}{l|r@{}lr@{}lr@{}lr@{}lr@{}l}
$h=1$  & LF && LS && L3 && LMV && LD &\\
  \hline
FC & 84.28 &$^{***}$& 0.9660 &$^{***}$& 1807 &$^{***}$& 0.7905 &$^{***}$& 8.1319 &$^{***}$\\
  ARMA & 85.09 &$^{**}$& 0.9950 &  & 1823 &$^{**}$& 0.7916 &  & 8.1533 &  \\
  ARFIMA & 86.22 &$^{**}$& 0.9955 &  & 1829 &$^{**}$& 0.7911 &$^{**}$& 8.1586 &  \\
  ARFIMA.chol & 87.82 &$^{**}$& 1.0830 &  & 1860 &$^{**}$& 0.7920 &$^{*}$& 8.1723 &$^{**}$\\
  CAW.diag & 85.97 &$^{**}$& 1.0254 &  & 1843 &$^{**}$& 0.7930 &  & 8.1869 &  \\
  CAW.dcc & 86.32 &$^{**}$& 1.0037 &  & 1866 &$^{**}$& 0.7928 &  & 8.3021 &  \\
  CAW.acomp & 85.77 &$^{**}$& 1.0268 &  & 1814 &$^{**}$& 0.7932 &  & 8.2482 &  \\
  CAW.mcomp & 90.72 &$^{**}$& 1.0301 &  & 1904 &$^{**}$& 0.7929 &  & 8.2478 &  \\
\end{tabular}
\caption{Out of sample risks for $h=1$ as described in section \ref{sec:oos}. In different rows, we consider the fractional 			components (FC) and several benchmark models, namely a diagonal vector ARMA(2,1) and a diagonal vector AR\-FI\-MA(1,$d$,1) model, the conditional autoregressive Wishart (CAW) model of \citet{GolGriLi2012}, a dynamic correlation specification (CAW-DCC) of \citet{BauStoVi2012}, and additive and multiplicative components Wishart models as proposed by \citet{JinMah2013}. Asterisks denote the best performing model ($^{***}$), models in the 80\% model confidence set ($^{**}$) and additional models in the 90\% model confidence set ($^{*}$). As loss functions, we consider the Frobenius norm (LF), the Stein norm (LS), the predictive densities (LD), the minimum-variance portfolio variance (LMV) and the L3-Loss (L3).} \label{tab:oos1}
\end{table}
\begin{table}[ht]
\centering
\begin{tabular}{l|r@{}lr@{}lr@{}lr@{}lr@{}l}
 $h=5$ & LF && LS && L3 && LMV && LD &\\
  \hline
FC & 135.28 &$^{***}$& 1.3766 &$^{***}$& 2463 &$^{***}$& 0.8011 &$^{***}$& 8.2490 &$^{***}$\\
  ARMA & 134.43 &$^{**}$& 1.4046 &  & 2498 &$^{**}$& 0.8025 &  & 8.2688 &  \\
  ARFIMA & 135.12 &$^{**}$& 1.3974 &  & 2492 &$^{**}$& 0.8015 &$^{**}$& 8.2664 &  \\
  ARFIMA.chol & 140.43 &$^{**}$& 1.5348 &  & 2557 &$^{**}$& 0.8021 &$^{*}$& 8.3113 &$^{**}$\\
  CAW.diag & 137.34 &$^{**}$& 1.4356 &  & 2612 &$^{**}$& 0.8038 &  & 8.3184 &  \\
  CAW.dcc & 137.77 &$^{**}$& 1.4094 &  & 2627 &$^{**}$& 0.8039 &  & 8.4150 &  \\
  CAW.acomp & 139.22 &$^{**}$& 1.4443 &  & 2558 &$^{**}$& 0.8028 &  & 8.3311 &  \\
  CAW.mcomp & 142.26 &$^{**}$& 1.4489 &  & 2590 &$^{**}$& 0.8028 &  & 8.3399 &  \\
\end{tabular}\caption{Out of sample risks for $h=5$ as described in section \ref{sec:oos}. For details on the abbreviations see table \ref{tab:oos1}.} \label{tab:oos2}
\end{table}
\begin{table}[ht]
\centering
\begin{tabular}{l|r@{}lr@{}lr@{}lr@{}lr@{}l}
$h=10$ & LF && LS && L3 && LMV && LD &\\
  \hline
FC & 170.07 &$^{**}$& 1.7033 &$^{**}$& 2837 &$^{**}$& 0.8102 &$^{**}$& 8.3118 &$^{***}$\\
  ARMA & 172.03 &$^{**}$& 1.6985 &$^{**}$& 2890 &$^{**}$& 0.8088 &$^{*}$& 8.3519 &  \\
  ARFIMA & 168.55 &$^{***}$& 1.6716 &$^{***}$& 2837 &$^{***}$& 0.8076 &$^{***}$& 8.3372 &  \\
  ARFIMA.chol & 173.43 &$^{**}$& 1.8455 &  & 2900 &$^{**}$& 0.8103 &$^{**}$& 8.3893 &  \\
  CAW.diag & 176.50 &$^{**}$& 1.7399 &$^{**}$& 2980 &$^{**}$& 0.8110 &$^{**}$& 8.4105 &  \\
  CAW.dcc & 178.11 &$^{**}$& 1.7120 &$^{**}$& 2986 &$^{**}$& 0.8101 &$^{**}$& 8.4973 &  \\
  CAW.acomp & 177.37 &$^{**}$& 1.7366 &$^{**}$& 2947 &$^{**}$& 0.8093 &$^{**}$& 8.4146 &  \\
  CAW.mcomp & 181.04 &$^{**}$& 1.7265 &$^{**}$& 3009 &$^{**}$& 0.8096 &$^{**}$& 8.4248 &  \\
\end{tabular}\caption{Out of sample risks for $h=10$ as described in section \ref{sec:oos}. For details on the abbreviations see table \ref{tab:oos1}.} \label{tab:oos3}
\end{table}
\begin{table}[ht]
\centering
\begin{tabular}{l|r@{}lr@{}lr@{}lr@{}lr@{}l}
$h=20$& LF && LS && L3 && LMV && LD &\\
  \hline
FC & 199.42 &$^{***}$& 2.0461 &$^{**}$& 3144 &$^{***}$& 0.8225 &$^{**}$& 8.3778 &$^{***}$\\
  ARMA & 208.07 &$^{**}$& 2.0980 &$^{**}$& 3231 &  & 0.8224 &$^{**}$& 8.4314 &  \\
  ARFIMA & 200.33 &$^{**}$& 2.0305 &$^{***}$& 3162 &$^{**}$& 0.8209 &$^{***}$& 8.4049 &  \\
  ARFIMA.chol & 203.22 &$^{**}$& 2.1910 &$^{**}$& 3201 &$^{**}$& 0.8219 &$^{**}$& 8.4738 &  \\
  CAW.diag & 214.60 &$^{*}$& 2.1580 &$^{**}$& 3326 &$^{*}$& 0.8241 &$^{**}$& 8.5034 &  \\
  CAW.dcc & 217.52 &  & 2.1698 &$^{**}$& 3331 &  & 0.8231 &$^{**}$& 8.6158 &  \\
  CAW.acomp & 211.33 &  & 2.1165 &$^{**}$& 3289 &$^{*}$& 0.8214 &$^{**}$& 8.5028 &  \\
  CAW.mcomp & 209.78 &$^{**}$& 2.0858 &$^{**}$& 3282 &$^{**}$& 0.8225 &$^{**}$& 8.5158 &  \\
\end{tabular}\caption{Out of sample risks for $h=20$ as described in section \ref{sec:oos}. For details on the abbreviations see table \ref{tab:oos1}.} \label{tab:oos4}
\end{table}

\clearpage


\end{document}